\documentclass[twocolumn]{aastex631}

\usepackage{graphicx}
\usepackage{amsmath} 
\usepackage{mathrsfs}
\usepackage{enumerate}
\usepackage{subfigure}

\shorttitle{Deep modeling of quasar disk reprocessing}
\shortauthors{Fagin et al.}

\begin{document}

\title{Joint Modeling of Quasar Variability and Accretion Disk Reprocessing Using Latent Stochastic Differential Equations}

\correspondingauthor{Joshua Fagin}
\email{jfagin@gradcenter.cuny.edu}

\author[0000-0001-8723-6136]{Joshua~Fagin}
\affiliation{The Graduate Center of the City University of New York, 365 Fifth Avenue, New York, NY 10016, USA}
\affiliation{Department of Astrophysics, American Museum of Natural History, Central Park West and 79th Street, NY 10024-5192, USA}
\affiliation{Department of Physics and Astronomy, Lehman College of the CUNY, Bronx, NY 10468, USA}

\author[0000-0001-8797-725X]{James~Hung-Hsu~Chan}
\affiliation{Department of Astrophysics, American Museum of Natural History, Central Park West and 79th Street, NY 10024-5192, USA}
\affiliation{Department of Physics and Astronomy, Lehman College of the CUNY, Bronx, NY 10468, USA}

\author[0009-0009-6932-6379]{Henry~Best}
\affiliation{The Graduate Center of the City University of New York, 365 Fifth Avenue, New York, NY 10016, USA}
\affiliation{Department of Astrophysics, American Museum of Natural History, Central Park West and 79th Street, NY 10024-5192, USA}
\affiliation{Department of Physics and Astronomy, Lehman College of the CUNY, Bronx, NY 10468, USA}
\affiliation{Department of Theoretical Physics and Astrophysics, Faculty of Science,
Masaryk University, Kotlářská 2, CZ-611 37 Brno, Czech Republic}

\author[0009-0000-4476-5003]{Matthew~O'Dowd}
\affiliation{The Graduate Center of the City University of New York, 365 Fifth Avenue, New York, NY 10016, USA}
\affiliation{Department of Astrophysics, American Museum of Natural History, Central Park West and 79th Street, NY 10024-5192, USA}
\affiliation{Department of Physics and Astronomy, Lehman College of the CUNY, Bronx, NY 10468, USA}

\author[0000-0002-5956-851X]{K.~E.~Saavik~Ford}
\affiliation{The Graduate Center of the City University of New York, 365 Fifth Avenue, New York, NY 10016, USA}
\affiliation{Department of Astrophysics, American Museum of Natural History, Central Park West and 79th Street, NY 10024-5192, USA}
\affiliation{Department of Science, CUNY Borough of Manhattan Community College, 199 Chambers St, New York, NY 10007, USA}
\affiliation{Flatiron Institute, 162 Fifth Avenue, New York, NY 10010, USA}

\author[0000-0002-3168-0139]{Matthew~J.~Graham}
\affiliation{Department of Physics, Maths and Astronomy, California Institute of Technology, 1200 E California Blvd, Pasadena, CA 91125, USA}

\author[0000-0002-0692-1092]{Ji~Won~Park}
\affiliation{SLAC National Accelerator Laboratory, Menlo Park, CA 94025, USA}
\affiliation{Prescient Design, Genentech, South San Francisco, CA 94080, USA}

\author[0000-0002-5814-4061]{V.~Ashley~Villar}
\affiliation{Center for Astrophysics \textbar{} Harvard \& Smithsonian, 60 Garden Street, Cambridge, MA 02138-1516, USA}

\begin{abstract}
Quasars are bright active galactic nuclei powered by the accretion of matter around supermassive black holes at the center of galaxies. Their stochastic brightness variability depends on the physical properties of the accretion disk and black hole. The upcoming Rubin Observatory Legacy Survey of Space and Time (LSST) is expected to observe tens of millions of quasars, so there is a need for efficient techniques like machine learning that can handle the large volume of data. Quasar variability is believed to be driven by an X-ray corona, which is reprocessed by the accretion disk and emitted as UV/optical variability. We are the first to introduce an auto-differentiable simulation of the accretion disk and reprocessing. We use the simulation as a direct component of our neural network to jointly model the driving variability and reprocessing, trained with supervised learning on simulated LSST-like 10 yr quasar light curves. We encode the light curves using a transformer encoder, and the driving variability is reconstructed using latent stochastic differential equations, a physically motivated generative deep learning method that can model continuous-time stochastic dynamics. By embedding the physical processes of the driving signal and reprocessing into our network, we achieve a model that is more robust and interpretable. We demonstrate that our model outperforms a Gaussian process regression baseline and can infer accretion disk parameters and time delays between wave bands, even for out-of-distribution driving signals. Our approach provides a powerful framework that can be adapted to solve other inverse problems in multivariate time series. 
\end{abstract}


\section{Introduction} 

Active galactic nuclei (AGN) are thought to be powered by the accretion of matter around supermassive black holes at the center of galaxies~\citep{Salpeter64,Zeldovich64}. Quasars are bright AGN with unobscured accretion disks, and are some of the brightest objects in the Universe. They are observable at extreme cosmological distances, making them powerful probes of the early Universe~\citep{Mortlock11, Banados18}. Quasars are also thought to play an important role in galaxy evolution~\citep{Franceschini_1999,Kauffmann_2000,Hoshi_2024}. The variability of quasar brightnesses has been studied since their discovery~\citep{1963Natur1,1963Natur2,1963ApJ,1963Natur3,1963Natur4}. These brightness variations are related to the physical properties of the black hole and accretion disk that power them. Furthermore, investigating the physics governing quasar light curves offers valuable insights into our understanding of cosmology~\citep{Khadka_2020,Czerny_2023}. 

The UV/optical variability of quasars is most often modeled as an X-ray driving variability source corona above the black hole that illuminates a geometrically thin, optically thick accretion disk~\citep{Shakura_1973, Cackett2007}. The reprocessing of the driving variability to the UV/optical emitting regions of the accretion disk is represented by the so-called transfer functions, and introduces wavelength-dependent time lags~\citep{Blandford82}, ranging from less than a day for small supermassive black holes ($\sim 10^7 M_\odot$) to several tens of days ($\sim 10^{10} M_\odot$). These time lags can be measured through continuum reverberation mapping of UV/optical light curves to probe the relative size scales of the emitted regions, and are interconnected to properties of the accretion disk and black hole~\citep{Cackett_2021,Jha_2022,Wang_2023}. 

Studies of quasar variability have found accretion disk sizes to be larger than the predictions of the standard thin-disk model~\citep{Shakura_1973} by a factor of $\sim$2--4~\citep{Mudd2018,Guo_2022,Jha_2022}. This is also consistent with accretion disk size measurements found using gravitational microlensing~\citep{Poindexter08, Poindexter10, Blackburne15, Munoz16, Morgan18}. New measurements and methods are needed to test accretion disk models and enhance our understanding of the physical processes governing quasar emissions.

Upcoming wide-field surveys such as the Rubin Observatory Legacy Survey of Space and Time (LSST) will observe an unprecedented quantity of data. The LSST main survey will cover 18,000~$\deg^2$ and is projected to monitor tens of millions of quasars over a 10 yr period with six UV/optical bandpass filters (\textit{ugrizy}) at 55--185 samplings per band, or around 800 total visits across the 10 yr out to a redshift of $z \sim 7.5$~\citep{abell2009lsst,Prsa_2023}. A smaller sky area of 200~$\deg^2$ known as the Deep Drilling Fields is expected to detect 40,000 additional ultrafaint AGN at higher cadences of about 1,000 samplings per band~\citep{brandt2018active}. Machine learning (ML) algorithms are well suited to analyze the vast amounts of data expected from LSST and other wide-field surveys. The quasar light curves from LSST pose challenges for traditional ML techniques due to being stochastic, multivariate, noisy, and irregularly sampled across bands with long seasonal gaps~\citep[see][for a review of ML methods on irregularly sampled time series]{shukla2021surveyprinciplesmodelsmethods}.

UV/optical variability light curves are most commonly modeled using Gaussian process regression (GPR). In this framework, quasar light curves are fit with a specific kernel, e.g., the kernel associated with the damped random walk (DRW) process~\citep{Zu_2013}. The optimized kernel parameters have been empirically shown to be related to properties of the quasar, such as the black hole mass~\citep{macleod2010modeling,Suberlak_2021}. 

The codebase \texttt{JAVELIN}~\citep{Zu_2011,Zu_2016,Fausnaugh_2016} uses a DRW kernel with top-hat transfer functions to simultaneously model the variability and time delays. The bluest band is taken as the \textit{effective} driving variability source and fit using the DRW, and the top-hat transfer functions are used to measure the time delays between the other bands. This method relies on Markov Chain Monte Carlo (MCMC) to sample the DRW kernel and transfer function parameters. By default, \texttt{JAVELIN} measures the time delays but does not directly extract the physical properties of the quasar. Simple parametric models can be fit to the time-delay measurements as a secondary step to measure the accretion disk size and sometimes the temperature profile, although it is most frequently fixed to the thin-disk case. In~\citet{Mudd2018}, \texttt{JAVELIN} was modified to measure the accretion disk size directly by fixing the time delays to the thin-disk model. 

The codebase \texttt{CREAM}~\citep{Starkey_2015} is similar to \texttt{JAVELIN} but uses the thin-disk transfer functions directly instead of top hats. Instead of treating the bluest band as the effective driving variability, \texttt{CREAM} explicitly reconstructs the driving variability by modeling it as a Fourier series. MCMC is used to optimize the Fourier components of the driving variability and the accretion disk parameters, and both the driving variability and transfer function kernels are reconstructed. In theory, \texttt{CREAM} can recover the product of the black hole mass and accretion rate, $M\dot{M}$, and the inclination angle, although the inclination cannot be recovered without very-high-fidelity data. For example, it was used in~\citet{Fausnaugh_2018} to model two Seyfert 1 galaxies to measure $M\dot{M}$ and to constrain the inclination for one galaxy.

Currently, \texttt{JAVELIN} and \texttt{CREAM} are the two most advanced methods of measuring time delays and accretion disk sizes in continuum reverberation mapping. However, both of them rely on computationally expensive MCMC sampling, which may be infeasible to apply to the tens of millions of quasar light curves expected from LSST. In addition, the DRW kernel in \texttt{JAVELIN} is only fit with respect to the bluest band; however, the information in each band is highly correlated to the other bands, so jointly modeling the light curves and time delays would improve the performance. This is especially the case for sparsely and irregularly sampled cadences like those expected from LSST. UV/optical light curves have also been shown to significantly deviate from a DRW~\citep{Yu_2022}, so a more flexible fitting method would be beneficial. \texttt{CREAM} does jointly model the driving variability and transfer functions; however, the driving variability is modeled as a Fourier series. This could be challenging for long light curves like what we expect from LSST, since the driving signal is expected to be stochastic across many time scales. Ideally, our methods would also be able to learn from features across an entire sample of light curves. For example, there is additional information in the mean brightnesses of each band that could be combined with time-delay measurements to better estimate accretion disk parameters and break some of the degeneracies. Our ML approach improves upon all these areas.

ML methods have been used to model quasar variability in several different ways. \citet{Tachibana_2020} used a recurrent auto-encoder to model quasar variability, and found it to perform better than a DRW model when applied to real data. \citet{Snchez_S_ez_2021} used a recurrent variational auto-encoder for anomaly detection to find changing-look AGN. \citet{JiWon_2021} introduced a method to simultaneously reconstruct quasar light curves and predict accretion disk parameters using attentive neural processes. \citet{Hajdinjak_2022} introduced conditional neural processes to model quasar variability. \citet{Sheng_2022} applied stochastic recurrent neural networks (RNNs) to reconstruct simulated LSST light curves. \citet{Danilov2022} developed a neural inference Gaussian processes method to fit quasar light curves. \citet{li2024fastflexibleinferenceframework} demonstrated how simulation-based inference could be used to predict accretion disk parameters. For microlensed quasar light curves, \citet{Vernardos_2019} used a convolutional neural network to measure the accretion disk size and temperature profile in simulated light curves, and~\citet{Best_2024} predicted the black hole mass, inclination angle, and impact angle. \citet{fagin2024predictinghighmagnificationevents} developed a method of predicting high-magnification microlensing events through real-time classification with simulated LSST light curves using an RNN. 

\citet{Fagin_2024} introduced latent stochastic differential equations (SDEs) as a method to reconstruct simulated LSST-like quasar light curves and simultaneously predict the accretion disk and variability parameters. Latent SDEs are a type of generative neural network that can model continuous-time stochastic dynamics~\citep{Torch_SDE}. They are physically motivated by the fact that quasar light curves are generally well described by SDEs such as the DRW or higher-order continuous-time autoregressive moving-average (CARMA) processes~\citep{Yu_2022}. Latent SDEs can be viewed as infinite-dimensional variational autoencoders~\citep{VAE,rezende2014stochastic} with an SDE-induced process as their latent state. \citet{Fagin_2024} found the deep learning method to be superior to a multitask GPR baseline in reconstructing LSST light curves. In addition, their model simultaneously performed parameter inference based on the context vector of the encoder and the latent vector of the SDE. They were able to predict the black hole mass, temperature slope, and inclination angle, as well as the parameters of the DRW driving variability signal.

While the latent SDE method of~\citet{Fagin_2024} is able to simultaneously reconstruct the light curve and perform parameter inference, the reconstructed light curve and parameter predictions do not necessarily correspond to the same time delays. In this work, we are the first to combine the light-curve reconstruction and parameter inference into a self-consistent, unified framework. This is achieved by developing the first auto-differentiable simulation of the accretion disk and including it into the architecture of our ML model. Within our NN, we use a latent SDE to generate the X-ray driving variability. We then predict the accretion disk parameters, which are converted to the corresponding transfer functions using our auto-differentiable simulation of the disk. The reconstructed driving variability is convolved with the transfer functions and then scaled to produce the mean best-fit reconstruction of each observed UV/optical band, along with their uncertainties. In addition, we predict the variability parameters of the driving signal and their uncertainties. The relative time delays between bands and their uncertainties are also predicted from the mean time delay of the reconstructed transfer functions. 

A recurrent inference machine~\citep[RIM;][]{putzky2017recurrent} is a technique that iteratively refines its predictions by feeding previous outputs back into the model. It has been used in several astrophysics applications~\citep{Morningstar_2019,modi2021cosmicrim,Adam_2023,rhea2023unraveling}. Our goal in this work is to solve the blind deconvolution inverse problem of recovering the driving signal and transfer functions given the observations of our light curve. This is made particularly challenging given the stochastic nature of quasar variability and the fact that our observations are irregular and sparsely sampled. We use RIM in our NN to iteratively improve upon the light-curve reconstruction and accretion disk parameter estimation. 

In Section~\ref{sec:sim}, we describe how we build a realistic simulation of quasar light curves including an auto-differentiable version that is incorporated into our ML model. In Section~\ref{sec:model}, we present our NN architecture and training. In Section~\ref{sec:results}, we give our results on our model's performance in light-curve reconstruction and parameter inference. In Section~\ref{sec:discussion}, we discuss the results of our ML model and future prospects, and Section~\ref{sec:conclusion} gives our concluding remarks. Throughout this work, we assume a flat $\Lambda$CDM cosmology with $H_0 = 70$ $\text{km}\, \text{s}^{-1}\, \text{Mpc}^{-1}$, $\Omega_{m} = 0.3$, and $\Omega_{\Lambda} = 0.7$.

\section{Light-curve Simulation} \label{sec:sim}

We train our ML model with realistic simulations of LSST 10 yr light curves. The reprocessing of the X-ray driving variability by the accretion disk to the UV/optical wavelength $\lambda$ is modeled by:  
\begin{equation} \label{eq:reprocessing}
F_\lambda(t, \lambda) = \bar{F}_\lambda(\lambda) + \Delta F_\lambda(\lambda)  \int_{0}^{\infty} X(t-\tau) \psi(\tau | \lambda) \text{d}\tau \, ,
\end{equation}
where $F_\lambda(t, \lambda)$ is the observed flux, $\bar{F}_\lambda(\lambda)$ is the mean flux, $\Delta F_\lambda(\lambda)$ is the amplitude of the variable flux, $X(t)$ is the normalized driving variability (mean zero and variance one), and $\psi(\tau | \lambda)$ is the transfer function kernel~\citep{Cackett2007,Starkey_2015,chan2024reverberationmappinglamppostwind}. Section~\ref{sec:driving_variability} describes how we model the driving variability $X(t)$. Section~\ref{sec:accretion_disk} introduces our auto-differentiable model of the accretion disk reprocessing and transfer functions $\psi(\tau | \lambda)$. In Section~\ref{sec:spectrum}, we construct realistic spectra from our accretion disk model using templates for the spectral lines, host-galaxy flux, and extinction. We then integrate the spectrum across the filter response functions of each LSST band to get the mean flux and variability amplitude of each band, $\bar{F}_\lambda(\lambda)$ and $\Delta F_\lambda(\lambda)$. Section~\ref{sec:parameter_range} gives the parameter ranges for building our training set, and Section~\ref{sec:LSST_observations} describes how the time series is degraded to mimic LSST observational cadences and noise.

\subsection{Quasar Driving Variability Model} \label{sec:driving_variability}

UV/optical quasar light curves are often modeled as a DRW, a type of Gaussian process also known as the Ornstein-Uhlenbeck process~\citep{williams2006gaussian,Zu_2013}. A DRW signal $X(t)$ is governed by the SDE:
\begin{equation} \label{eq:DRW}
dX(t) = -\frac{1}{\tau} X(t)\ dt +\sigma \sqrt{dt}\ \epsilon(t) + b\ dt
\end{equation}
where $\epsilon(t)$ is a white noise process with a mean of zero and variance of one, $\tau$ is the characteristic timescale, $b$ is related to the mean of the process $\overline{X}=b\tau$, and $\sigma$ is related to the standard deviation defined by the asymptotic structure function $\text{SF}_\infty = \sigma\sqrt{\tau/2}$~\citep{Kelly2009}. The DRW can alternatively be characterized by its power spectral density (PSD), given by:
\begin{equation} \label{eq:DRW_PSD}
P(\nu) = \frac{4\tau\text{SF}_\infty^2}{1+(2\pi \tau \nu)^2} \, ,
\end{equation}
for frequency $\nu$. This model is useful because GPR can be used to measure the variability parameters $\tau$ and $\text{SF}_\infty$ using the kernel of the Gaussian process:
\begin{equation} \label{eq:DRW_kernel}
k(\Delta t) = \text{SF}_\infty^2 \, \exp\left(-\frac{\Delta t}{\tau} \right) 
\end{equation}
where $\Delta t$ is the time separation of two observations. The measured values of $\tau$ and $\text{SF}_\infty$ have been empirically shown to relate to properties of the accretion disk such as the black hole mass~\citep{macleod2010modeling,Suberlak_2021}. More complex Gaussian processes have also been used, such as higher-order CARMA processes~\citep{Yu_2022}. These CARMA processes are the solution to higher-order SDEs than the DRW given in Equation~(\ref{eq:DRW}).

The X-ray driving variability has been empirically shown to be better modeled by a broken power-law (BPL) PSD than a DRW. We generate our simulated driving signal from a bended BPL PSD given by:
\begin{equation} \label{eq:bended_broken_power_law}
P(\nu) \propto
\nu^{-\alpha_L}\left(1+\left(\frac{\nu}{\nu_b}\right)^{\alpha_H-\alpha_L}\right)^{-1} \, ,
\end{equation}
where $\nu_b$ is the break frequency between the lower power-law slope $\alpha_L$ and the higher power-law slope $\alpha_H$. This PSD implies that $P(\nu) \propto \nu^{-\alpha_L}$ at low frequencies when $f \ll \nu_b$ and $P(\nu) \propto \nu^{-\alpha_H}$ when $f \gg \nu_b$. A form of the bended BPL PSD has been extensively used to model the X-ray variability~\citep[e.g.,][]{McHardy_2004,Uttley_2005,Oneill_2005,soton466440,Markowitz_2010,smith2018kepler,Sartori_2019,Yang_2022,yuk2023correlation,Czerny_2023}. Although the bended BPL is not a Gaussian process, its parameters can be measured by directly fitting the PSD. In the special case where $\alpha_L = 0$, $\alpha_H = 2$, and $\nu_b = 1/(2\pi\tau)$, the bended BPL recovers the PSD of the DRW in Equation~(\ref{eq:DRW_PSD}). Typically, the BPL for X-ray variability has been measured with $\alpha_L \sim 1$ and $\alpha_H \sim 3$. Similarly to the DRW, the parameters of the BPL have been shown to relate to properties of the accretion disk and black hole~\citep{McHardy_2005,arevalo2023universalpowerspectrumquasars}. For example, the break frequency decreases with an increased black hole mass. In this work, we do not correlate the variability parameters to the accretion disk parameters, since we are interested in measuring these correlations in data without bias.

We can generate light curves from any PSD using the method of~\citet{Timmer_1995} by taking the inverse Fourier transform with randomized complex phases. This method has been used to generate X-ray variability with the bended BPL~\citep[e.g.,][]{Czerny_2023}. We generate the driving variability at a length 8 times larger than our desired signal, and keep only the initial desired length. This is to avoid the boundary conditions of the Fourier transform, and so we can set the asymptotic mean and standard deviation to the longer time series without bias (described in more detail in Section~\ref{sec:spectrum}). Generating the driving signal with 8 times the target length should be more than sufficient to achieve these goals. The driving variability is generated at daily intervals and in magnitude, so the flux is always positive.

The power spectrum in Equation~(\ref{eq:bended_broken_power_law}) is in the rest frame of the quasar. We predict the break frequency in the observer frame to avoid degeneracies with the redshift. The posteriors of our break frequency and redshift can be combined after the fact to obtain the rest-frame break frequency: $\nu_b^{\text{rest}} = (1+z)\nu_b^{\text{obs}}$.

\subsection{Auto-Differentiable Accretion Disk Model} \label{sec:accretion_disk}

Our goal is to use the accretion disk model to simulate our training set and as a direct component of our NN. We implement the accretion disk and transfer functions in \texttt{PyTorch}~\citep{Pytorch}, enabling automatic differentiation and allowing us to use the model with gradient-based optimization (i.e., we can use backpropagation and gradient descent to update the weights of our NN). The transfer functions $\psi(\tau | \lambda, \boldsymbol{\eta})$ are parameterized by a set of accretion disk and black hole parameters, given by the vector~$\boldsymbol{\eta}$.

We use a modified version of the Novikov-Thorne (NT) model~\citep{Novikov_1973}, the relativistic version of the Shakura-Sunyaev (SS) thin-disk model~\citep{Shakura_1973}. At large radii, the NT and thin-disk models predict viscous temperature slopes of \mbox{$T_{\text{visc}} \propto R^{-3/4}$}. Quasar microlensing studies have measured a general temperature profile of the form \mbox{$T \propto R^{-\beta}$} and favor shallower slopes $\beta < 3/4$~\citep{Cornachione_2020}. Some accretion disk models predict shallower or steeper slopes than the thin-disk slope of $\beta = 3/4$. For example, the slim-disk model predicts a shallower slope of $\beta \approx 0.5$~\citep{SlimDisk} while the magnetorotational
instability model of~\citet{Agol_2000} predicts a steeper slope of $\beta = 7/8$. The temperature slope can also be modified due to the presence of wind outflows, where the accretion rate becomes radially dependent~\citep{Blandford_1999,You_2016, Li_2018, Sun_2018,huang2023black,chan2024reverberationmappinglamppostwind}. We use the NT plus lamppost temperature profile:
\begin{align} \label{eq:temperature}
T_{\text{eff}}^4 &= T_{\text{visc}}^4+T_{\text{lamp}}^4 \nonumber\\
 &= \frac{3GM\dot{M}(R)}{8\pi R_g^3 \sigma_{\text{SB}}} f_{\text{NT}}(R, a)   \nonumber \\
 &+ \frac{(1-A)\eta_X M\dot{M}_0 c^2}{2\pi\sigma_{\text{SB}}}\frac{H}{(R^2+H^2)^{3/2}} \, , 
\end{align}
where $R$ is the radius away from the black hole on the disk, $M$ is the black hole mass, $\dot{M}(R)$ is the radially dependent accretion rate, $f_{\text{NT}}$(R, a) is the dimensionless flux factor from the NT model given in Appendix~\ref{sec:appendix_NT_model}, $H$ is the corona height, $\eta_X$ is the X-ray radiative efficiency, and $\sigma_{\text{SB}}$ is the Stefan-Boltzmann constant. We could instead use the SS model with $f_{\text{SS}}$, but we choose to use the NT model since it includes general relativistic (GR) corrections. To model the deviations of the viscous temperature due to wind inflows or other deviations from the NT model, we take the accretion rate to be a power law of the form:
\begin{equation} \label{eq:wind}
\dot{M}(R) = \dot{M}_{\text{in}} \left(\frac{R}{R_\text{in}}\right)^s
\end{equation}
where $s$ is the power-law slope of the accretion rate related to the asymptotic slope of the viscous temperature profile $\beta$ by $s = 3-4\beta$, $R_\text{in}$ is the inner radius of the disk, equivalent to the innermost stable circular orbit (ISCO), and $\dot{M}_{\text{in}}$ is the accretion rate at $R_\text{in}$~\citep{chan2024reverberationmappinglamppostwind}. For reference, $R_{\text{in}} / R_g$ = $9, 6, 1$ for $a = -1, 0, 1$ respectively, where $R_g = GM/c^2$ is the gravitational radius~\citep{Novikov_1973}. 

We want to define the accretion rate based on the Eddington ratio. To do this, we set the accretion rate for the thin-disk case to \mbox{$\dot{M}_{\text{in}}(s = 0) = \dot{M}_{0}$} where \mbox{$\dot{M}_0 = \lambda_{\text{Edd}}\dot{M}_\text{Edd}$} for Eddington ratio $\lambda_{\text{Edd}}$ and Eddington accretion rate \mbox{$\dot{M}_\text{Edd} = 4\pi GMm_\text{p}/(\eta\sigma_\text{T}c^2$)}, where $m_\text{p}$ is the proton mass, $\sigma_\text{T}$ is the Thomson scattering cross section, and $\eta$ is the overall radiative efficiency factor. We can then set $\dot{M}_{\text{in}}$ such that the total viscous bolometric luminosity is fixed to the thin-disk case by:
\begin{equation} \label{eq:M_in}
\dot{M}_{\text{in}} = \dot{M}_{0} \frac{\int_{R_\text{in}}^\infty F_{\text{NT}}(R, a) R dR}{\int_{R_\text{in}}^\infty F_{\text{NT}}(R, a) \left(\frac{R}{R_{\text{in}}}\right)^s R dR}  \, ,
\end{equation}
where $F_{\text{NT}} = \sigma_{\text{SB}}T_{\text{visc}}^4$ is the flux of the NT model at $s = 0$. The radiative efficiency $\eta$ is set such that the total luminosity is constant with $a$ and is given by:
\begin{equation} \label{eq:eta}
    \eta = \sqrt{1-\left(1-\frac{2 R_g}{3 R_{\text{in}}}\right)} \, ,
\end{equation} 
in the NT model. For reference, $\eta = 0.0377, 0.0572, 0.4226$ for spin $a = -1, 0, 1$ respectively. We note that for the SS model, the radiative efficiency should instead be set to \mbox{$\eta = R_g/2 R_{\text{in}}$}. When $\beta < 0.5$, the bolometric luminosity diverges when integrating out to infinity (i.e., the denominator in Equation~(\ref{eq:M_in}) becomes very large), but here we only consider the case when \mbox{$\beta \in [0.5,1.0]$}. 

\begin{figure*}
    \centering    
    \includegraphics[width=0.97\textwidth]{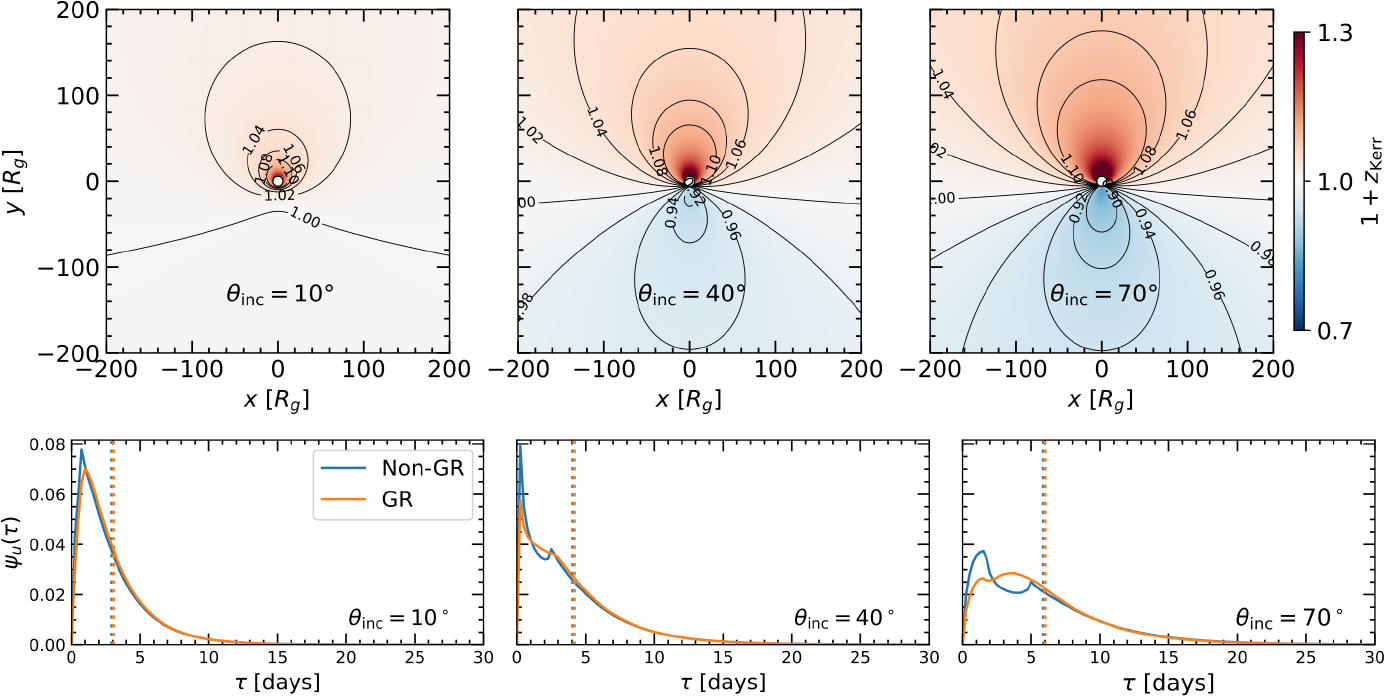}
    \caption{Here we demonstrate the impact of GR effects. The top panels show example redshift maps (from Equation~(\ref{eq:redshift_kerr})) with Keplerian angular velocity (from Equation~(\ref{eq:kepler_vel})) for three different inclination angles ($10^{\circ}$, $40^{\circ}$, and $70^{\circ}$ from left to right). The black hole spin to set to $a = 0$. The bottom panels show example transfer functions for the $u$ band without GR effects (blue) compared to the case where we include GR effects (orange) for the same inclination angles. The mean time delays are given by the vertical dotted lines. For the transfer functions, there are the additional parameters: $\log_{10}(M/M_\odot) = 9$, $\beta = 0.75$, $\lambda_{\text{Edd}} = 0.1$, $H = 16 R_g$, $z = 4$, $f_{\text{lamp}}=0.005$. The GR effects in the transfer functions include using the NT instead of the SS temperature profile and the gravitational redshifting and Doppler shifting due to the Kerr black hole (shown in top panels).}
    \label{fig:TF_compare_GR}
\end{figure*}

We include GR gravitational redshifting and Doppler shifting due to the Kerr black hole. In the rest frame of the disk, the effective temperature of each region of the disk is converted to disk flux through blackbody radiation:
\begin{equation}
    \label{eq:planck}
    B(\lambda_{\text{emit}}, T_{\text{eff}}) = \frac{2hc^{2}}{\lambda_e^{5}} \left( \frac{1}{e^{hc/(\lambda_{\text{emit}} k_{\text{B}} T_{\text{eff}})} - 1} \right) g^4, 
\end{equation}
where $h$ is the Planck constant, $k_\text{B}$ is the Boltzmann constant, $\lambda_{\text{emit}}$ is the emitted photon wavelength, and $g = 1/ (1+z_{\text{Kerr}})$ is the redshift factor due to both gravitational redshifting and relativistic Doppler beaming effects near the Kerr black hole~\citep{Cunningham73}. When simulating our transfer functions, we want to use the photons emitted at a redshift $\lambda_{\text{emit}}$ that will correspond to the observed wavelengths $\lambda_{\text{obs}}$, in our case the effective wavelength of each LSST wave band. We must therefore account for the cosmic redshifting of the photon as well as the gravitational redshift and relativistic Doppler shifting from the Kerr black hole. An emitted photon will be observed at a wavelength:
\begin{equation}
\lambda_{\text{emit}} = \frac{\lambda_{\text{obs}}}{(1+z_{\text{Kerr}})(1+z)} \, ,
\end{equation}
where $z$ is the cosmic redshift. The redshifting due to the Kerr black hole geometry is given by:
\begin{equation} \label{eq:redshift_kerr}
1+z_{\text{Kerr}} = \frac{(1+\Omega_K) r \sin(\theta) \sin(\phi)}{\sqrt{-g_{tt}-2 g_{t\phi} \Omega_K - \Omega_K^2 g_{\phi\phi}}} \, ,
\end{equation}
where each $g_{\mu\nu}$ are the metric components of a Kerr space-time, $\Omega_K$ is the angular velocity, $\theta = \theta_{\text{inc}}$ is the inclination angle of the disk, and $\phi$ is the azimuthal angle~\citep{Luminet_1079,Bhattacharyya_2001,Mastroserio_2018,Heydari_Fard_2023}. We approximate this effect without the need for GR ray tracing by ignoring the effects of light bending and assuming the accretion disk orbits the black hole at Keplerian angular velocity:
\begin{equation} \label{eq:kepler_vel}
\Omega_K = \frac{c/R_g}{(R/R_g)^{3/2}+a} \, ,
\end{equation}
where the disk rotates in a circular orbit around the black hole~\citep{Abramowicz_2013,Mastroserio_2018}. In this way, we can efficiently implement GR effects into our auto-differentiable accretion disk simulation. An example redshift map for different disk orientations is shown in the top panels of Figure~\ref{fig:TF_compare_GR}. The spin has very little effect on the redshift map except in the inner region of the disk near the ISCO. The bottom panels show example transfer functions with and without all the GR effects. At typical rest-frame wavelengths and low inclination, these effects have minimal impact on the transfer functions. However, at high redshifts and large inclination, where the shorter wavelengths allow us to probe the inner regions of the disk, these effects can alter the shape of the transfer function, although the mean time remains nearly unchanged. 

From the temperature profile defined in Equation~(\ref{eq:temperature}), the blackbody flux in Equation~(\ref{eq:planck}), and standard time lags of the lamppost thin-disk model, we can generate the transfer functions~\citep{Cackett2007}. See~\citet{chan2024reverberationmappinglamppostwind} for a full derivation of the transfer function. In order to be computationally efficient, we want to minimize the number of pixels we need in the grid to accurately calculate the transfer functions. We define an effective radius of the accretion disk by numerically solving for $k_\text{B} T_{\text{eff}}(R_c) = hc/\lambda$, where we use the wavelength of the reddest band (in our case the $y$ band). To account for cases where there is no solution, we define a characteristic radius by:
\begin{equation}
R_c = \operatorname*{argmin}_{R} \Big( \big| k_\text{B} T_{\text{eff}}(R) - hc/\lambda \big| \Big) \, ,
\end{equation}
and calculate the transfer function using a $1000\times 1000$ grid out to $100 R_c$. This effectively calculates the disk out to an infinite outer radius, since the flux decays to an insignificant amount well before $100 R_c$. The characteristic radius is the same as in~\citet{chan2024reverberationmappinglamppostwind} but using the full accretion disk model so cannot be solved for analytically. 

The transfer functions introduce a wavelength-dependent time lag:
\begin{equation} \label{eq:time_lag2}
    \bar{\tau}_{\lambda} = \frac{\int_{0}^{\infty} \psi(\tau | \lambda) \tau d\tau}{\int_{0}^{\infty} \psi(\tau | \lambda) d\tau} .
\end{equation}
We normalize the transfer functions to represent a probability distribution such that: $\int_{0}^{\infty} \psi(\tau | \lambda) d\tau = 1$ for all~$\lambda$. The transfer functions are calculated out to 800 days, chosen to be sufficiently long such that going out further is insignificant even for the largest-mass black holes. We evaluate the transfer functions on daily intervals. The influence that each parameter has on the mean time delays and standard deviation of the transfer functions can be found in Appendix~\ref{sec:appendix_parameters}.

\subsection{Spectrum Model} \label{sec:spectrum}

In addition to the transfer functions, we model the quasar spectrum to obtain the mean flux in each broadband filter consistently with the accretion disk parameters. We do not need to model the spectrum in the auto-differentiable simulation, since the mean brightnesses of each wave band are independent free parameters fit to the observations, rather than being fixed by the accretion disk parameters. 

\begin{figure*}
    \centering
    \includegraphics[width=0.97\textwidth]{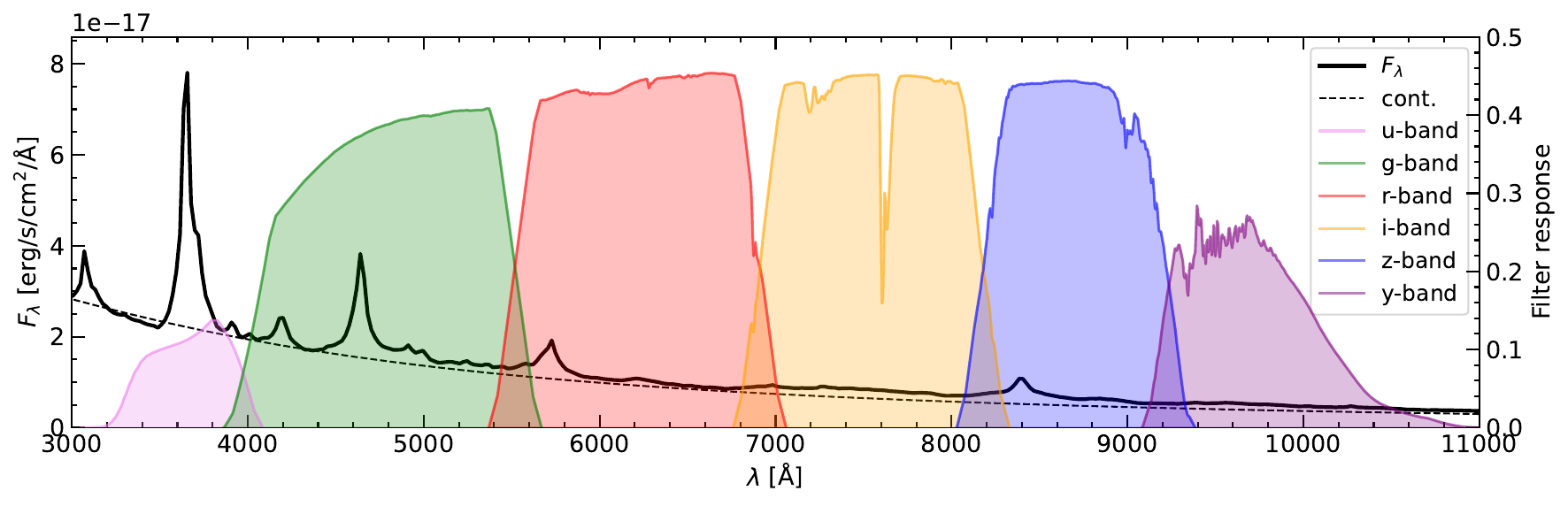}
    \caption{Example simulated quasar spectrum (black solid line) using the continuum of our accretion disk model (black dotted line) and templates from~\citet{Temple_2021}. The filter responses of the six LSST bands are included from~\citet{speclite}. We can integrate the quasar spectrum across the filter responses to obtain the observed flux in each filter. The parameters of this example spectrum are: $\log_{10}(M/M_\odot) = 8$, $\beta = 0.75$, $\theta_{\text{inc}} = 45^{\circ}$, $\lambda_{\text{Edd}} = 0.1$, $a = 0$, $H = 16 R_g$, $z = 2$, $f_{\text{lamp}}=0.005$, and $E(B-V) = 0.0$. At this redshift, the host-galaxy contribution is very minor.}
    \label{fig:spectrum}
\end{figure*}

To model the mean brightness in each band, we generate the quasar spectrum and then integrate it across the response function of each bandpass filter. Our accretion disk model that we defined in Section~\ref{sec:accretion_disk} can evaluate the mean flux of the disk at each wavelength for a given luminosity distance. We evaluate the mean fluxes 
corresponding to the observed wavelengths of the LSST filters (3,000--11,000\AA) to generate the continuum spectrum. To properly model the quasar spectrum, however, we need to also include the emission lines, flux from the host galaxy, spectral reddening due to extinction from dust along the line of sight, contamination in the Lyman-alpha forest, and the Lyman limit. These effects have been previously modeled by~\citet{Temple_2021} and calibrated for redshifts $0 < z < 5$. We use the spectrum model of~\citet{Temple_2021}, but replace the continuum from a BPL with the continuum generated from our accretion disk model. They use a set of templates to include the spectral bands, host-galaxy flux, and extinction. The Lyman limit is modeled by setting the flux at a rest-frame wavelength of $\lambda < 912 \text{\AA}$ to zero. Once the full spectrum is obtained, we use~\texttt{speclite}~\citep{speclite} to integrate the flux across the LSST broadband filters and obtain the mean magnitude of each wave band. Figure~\ref{fig:spectrum} shows an example simulated quasar spectrum with the LSST response functions. 

We first calculate the absolute magnitude of the $i$ band $M_i$ from the continuum, which is used to set the scaling of the emission lines and host galaxy. We add an additional scatter to our spectrum by adding $\mathcal{N}(0, 0.1)$ mag to $M_i$. The other main parameter is the extinction strength $E(B-V)$, which we randomly draw from $\mathcal{N}(0,0.075)$ and take its absolute value. The extinction has a significant effect on the quasar brightness. We further add additional scatter by varying the emission lines by controlling the relative scale height of the H$\alpha$, Ly$\alpha$, and near-line region, which we vary $\mathcal{N}(0,0.05)$. In real quasar data, individual emission lines can further vary in width and strength depending on the accretion disk parameters and the broad-line region of the disk. For example, if there is a stronger than average emission line in a waveband, then the quasar would appear brighter, and there would be longer time delays, so we may overestimate the black hole mass. We include an additional scatter in the mean brightness of each band drawn from $\mathcal{N}(0,0.025)$ mag. We also include an additional constant offset to all bands drawn from $\mathcal{N}(0,0.025)$ mag. 

We limit the redshift to the range $z \in [0.1, 5.0]$, which includes the vast majority of quasars that will be observed by LSST. Within our selected redshift range, due to the cutoff in the Lyman limit, we do not always observe across all bands. The bluest band of LSST starts at $3000 \text{\AA}$, so if the Lyman limit cutoff is redshifted past this range (starting at $z = 2.29$), parts of the LSST observations will be affected. At the highest redshifts, both the $u$ and $g$ bands will not be observed at all. To keep within the observational range of LSST, if the mean magnitude of the $i$ band is greater than 27 mag or less than 13 mag, then we resimulate a new quasar with a random set of parameters. The 27 magnitude limit represents the faintest possible objects observable by LSST, and the 13 magnitude lower limit accounts for saturation with bright sources. 

We obtain the mean brightness of each band both with and without the host galaxy, because the host-galaxy flux should not be variable. The driving signal is first produced in magnitude with mean zero and standard deviation $\sigma$, and added to the mean magnitude of each band without the host-galaxy contribution. Next, we convert from magnitude to flux and convolve with each transfer function kernel. The flux from the host galaxy is then added, and finally, we convert back from flux to magnitude.

As mentioned in Section~\ref{sec:driving_variability}, the driving signal is initially generated 8 times larger than needed to asymptotically set the mean and standard deviation in the longer signal. We then take the first eighth of the signal, where the local mean and standard deviation will not be the same as their asymptotic values. When constructing the training set, the driving signal is set to the normalization of the reference $i$ band's brightness before convolving with the transfer functions, since its mean brightness is arbitrary.

\subsection{Parameter Ranges} \label{sec:parameter_range}

The parameters to generate our light curves are given in Table~\ref{table:Parameter_table}. The light curves are simulated with 12 physical parameters, four for the driving variability and eight for the accretion disk. The driving variability depends on its standard deviation $\sigma$, the break frequency $\nu_b$, and the low- and high-frequency power laws $\alpha_L$ and $\alpha_H$. The black hole geometry is determined by its mass $M$ and dimensionless black hole spin $a$. The quasar is at an inclination angle $\theta_{\text{inc}}$ (with $0^{\circ}$ being face-on) and a redshift $z$. We also vary the corona height $H$, corona X-ray strength $f_{\text{lamp}}$, asymptotic viscous temperature slope $\beta$, and Eddington ratio $\lambda_{\text{Edd}}$. When sampling the light curves, we draw each parameter uniformly from its minimum and maximum range. There are also additional parameters related to the quasar spectrum that influence the mean magnitudes, in particular the extinction strength \mbox{$E(B-V)$}, that are not predicted by our ML model.

\begin{table}
\centering
\caption{Parameters and ranges to create a mock light curve and predicted by our ML model. The parameters are sampled uniformly between the minimum and maximum given in the table. The top four parameters are for the X-ray driving variability, while the bottom eight parameters relate to the black hole and accretion disk reprocessing.} 
 \begin{tabular}{c c c c} 
Parameter & Description & Min. & Max.  \\
 \hline\hline
$\sigma/\text{mag}$ & Variability amplitude & 0 & 0.5 \\
$\log_{10}(\nu_b/\text{day}^{-1})$ & Break frequency & $-3.5$ & 0.0 \\
$\alpha_L$ & Lower power law & 0.25 & 1.5 \\
$\alpha_H-\alpha_L$ & Higher power law & 0.75 & 2.75 \\
\hline
$\log_{10}(M/M_\odot)$ & Black hole mass & 7 & 10 \\
$a$ & Dimensionless spin & $-1$ & 1 \\
$\theta_{\rm inc}$ & Inclination angle & $0^{\circ}$ & $70^{\circ}$ \\
$(H-R_{\text{in}})/R_g$ & Corona height & 0 & 50 \\
$f_{\text{lamp}}$ & Lamppost strength & 0.002 & 0.007 \\
$\beta$ & Temperature slope & 0.3 & 1.0 \\
$z$ & Redshift & 0.1 & 5 \\
$\log_{10}(\lambda_{\text{Edd}})$ & Eddington ratio & $-2$ & 0 \\
 \hline\hline  
\end{tabular}
\label{table:Parameter_table}
\end{table}

We parameterize the ISCO height as $(H-R_{\text{in}})/R_g$, since the corona must always be above the ISCO. We parameterize $\alpha_H-\alpha_L$ instead of using $\alpha_H$ directly, because we expect $\alpha_H \ge \alpha_L$. The X-ray radiative efficiency factor \mbox{$\eta_X = (\eta/\lambda_{\text{Edd}}) (L_X/L_{\text{Edd}})$} where $L_X/L_{\text{Edd}} \sim 0.005$~\citep{Ursini_2020}. We parameterize the lamppost strength, which we define as $f_{\text{lamp}} = (1-A) (L_X/L_{\text{Edd}})$ since the albedo and X-ray luminosity ratio are degenerate. The albedo ranges from $A = 0$ for full absorption to $A = 1$ full
reflection, with typical values $A\sim 0.1$--$0.2$~\citep{Ursini_2020}. The driving variability parameters are chosen to be consistent with measurements from~\citet{soton466440}. 

\subsection{Mock LSST Observations} \label{sec:LSST_observations}

After the UV/optical light curves are simulated, we degrade them to mimic LSST-like observing cadences and noise in the same way as section 2.4 of~\citet{Fagin_2024}. The errors at each LSST observation are determined by:
\begin{equation} \label{eq:noise}
\sigma_{\text{LSST}}^2 = \sigma_{\text{sys}}^2 + \sigma_{\text{rand}}^2 \, ,
\end{equation}
where $\sigma_{\text{sys}}$ is the systematic error, and $\sigma_{\text{rand}}$ is the photometric noise. The systematic error is set to $0.005$ mag, the maximum value expected for LSST~\citep{LSST,Suberlak_2021}. The photometric noise depends on the brightness of each observation, and is expected to follow:
\begin{gather}
\sigma_{\text{rand}}^2 = (0.04-\gamma)x+\gamma x^2 \quad (\text{mag}^2) \, , \\
x = 10^{0.4(m-m_5)} \nonumber
\end{gather}
where $\gamma$ is a band-dependent factor, $m$ is the magnitude of each observation, and $m_5$ is the $5\sigma$ depth of a point source observed at the zenith of the observation. We expect $\gamma_{u} = 0.038$ and $\gamma_{g, r, i, z, y} = 0.039$~\citep{LSST,Sheng_2022}. We simulate LSST-like observations using \texttt{rubin\_sim}\footnote{\href{https://github.com/lsst/rubin\_sim}{https://github.com/lsst/rubin\_sim}} with the \texttt{baseline\_v2.1\_10yrs} rolling cadence, which gives the time and $m_5$ of each observation. We produce a random sample of 100,000 LSST-like observations by sampling anywhere in the sky with 750--1000 total observations across all bands. This filters the light-curve sample to include only the Wide Fast Deep observations, the main LSST survey~\citep[see Figure 1 of][]{Prsa_2023}. 

We randomly deviate each observation with variance $\sigma_{\text{LSST}}^2$. We combine observations to the nearest 1 day interval. We simulate 10.5 yr light curves with 10 yr of LSST observations, so our model is trained to reconstruct some time before and after the survey. The start time of the first LSST observation is randomly selected within the extra half-year. Any observation outside the range of 13--27 mag is excluded, which should be similar to the best-case magnitude limits of LSST. We discard observations in each band with fewer than 15 total observations across the 10 yr to account for bands near the magnitude limits.

\section{Machine learning model} \label{sec:model}

\subsection{Overview}

The overall goal of our ML model is to solve the inverse problem in Equation~(\ref{eq:reprocessing}) related to reconstructing the driving variability $X(t)$ and transfer function kernels $\psi(\tau | \lambda)$ given a set of noisy and sparsely sampled observations. In the context of quasar variability, this inverse problem involves deducing the underlying physical processes (e.g., the driving variability and accretion disk reprocessing physics) that produce the observed light curves. Unlike many inverse problems, where the forward process is well understood, this task is complicated by the fact that quasar variability itself is stochastic. This adds an additional layer of uncertainty to the problem, as the driving variability $X(t)$ is not directly observable with UV/optical data and must be inferred by the ML model.

Since the reprocessing is formulated as a convolution of the driving variability and transfer function, we are essentially training a ML model to solve a blind deconvolution problem. Blind deconvolution is a classic problem in signal processing where both the input signal (in this case, $X(t)$) and the kernel (here, $\psi(\tau | \lambda)$) must be inferred simultaneously from the observed data. This problem is particularly challenging because of the degeneracy between $X(t)$ and $\psi(\tau | \lambda)$, i.e. different combinations of driving variability and transfer functions can produce the same observed flux.  Moreover, the stochastic nature of $X(t)$ exacerbates this challenge, as it introduces variability that cannot be precisely predicted.

In general, recovering the transfer function kernels is an intractable problem because of this degeneracy. We could instead treat the bluest band as an effective driving signal and define effective kernels with respect to it. This is what is done with \texttt{JAVELIN} for example, but then we sacrifice most of the physics of the reprocessing on the disk. To make the problem tractable, we parameterize the transfer functions through our auto-differentiable simulation $\psi(\tau | \lambda, \boldsymbol{\eta})$ where $\boldsymbol{\eta}$ is the vector of accretion disk parameters, given in Table~\ref{table:Parameter_table}. The driving signal is reconstructed using a latent SDE~\citep{Torch_SDE,Fagin_2024}, parameterized by the context at each time and the latent vector $\boldsymbol{\hat{z}}$. We then solve for the flux by embedding the physics of the reprocessing of the driving variability into our NN architecture by:
\begin{equation} \label{eq:flux_reconstruction}
\underbrace{F_\lambda(t, \lambda|\boldsymbol{\hat{z}}, \boldsymbol{\hat{\eta}})}_{\substack{{\rm flux} \\ {\rm predicted}}} \propto \int_{0}^{\infty} \underbrace{X(t-\tau | \boldsymbol{\hat{z}})}_{\substack{{\rm latent} \\ {\rm SDE}}} \underbrace{\psi(\tau | \lambda,\boldsymbol{\hat{\eta}})}_{\substack{{\rm auto{\rm -}diff} \\ {\rm transfer\;function}}} \text{d}\tau \, ,
\end{equation}
where this convolution of the driving variability and transfer functions is evaluated numerically within our ML model using fixed-grid numerical integration. Our ML model also quantifies the uncertainty in the reconstructed $F_\lambda(t, \lambda)$, $X(t)$, $\boldsymbol{\hat{\eta}}$, the variability parameters, and the mean time delays $\bar{\tau}_\lambda$ coming from each reconstructed transfer function $\psi(\tau | \lambda, \boldsymbol{\hat{\eta}})$. 

The input to the ML model is the brightness and error values (both in magnitude), for a total of 12 features at each time step (six bands and six errors). For training stability, each band is normalized to have mean zero and standard deviation one, but the mean and standard deviation are used to predict the latent space of the SDE and the parameter posteriors. The reconstructed driving signal and UV/optical variability are unnormalized after they are generated by the ML model. For each band that is not observed at a given time step, we set both the brightness and error to a dummy value to be masked by the ML model. The outputs of our ML model are: the best-fit reconstruction of the UV/optical light curves, the reconstructed driving signal, the reconstructed transfer functions, the predicted accretion disk and driving variability parameters and uncertainty, and the predicted relative time delays between bands and uncertainties (where the $i$ band is arbitrarily chosen as the reference band). The model is trained using supervised learning to best reconstruct the light curve, driving signal, parameters, and time delays.

\subsection{Proof-of-concept Recurrent Inference Machine} \label{sec:RIM}

We use RIM to compare the reconstruction of our light curve with the observations, and iteratively adjust the accretion disk parameters by \mbox{$\boldsymbol{\hat{\eta}}_{i+1} =\boldsymbol{\hat{\eta}}_i + \Delta\boldsymbol{\hat{\eta}}_i$} and the latent space of the latent SDE by \mbox{$\boldsymbol{\hat{z}}_i = \boldsymbol{\hat{z}}_i + \Delta\boldsymbol{\hat{z}}_i$} in Equation~(\ref{eq:flux_reconstruction}) with iteration $i$ and initial values of zero. Due to constraints in GPU memory, only a small batch size can be used, and this process can be computationally costly. We therefore train the ML model first without this process and then test training using RIM with three iterations, although ideally, we would use more iterations and train longer.

\subsection{Model Architecture} \label{sec:model_architecture}

\begin{figure*}
    \centering
    \includegraphics[width=0.85\textwidth]{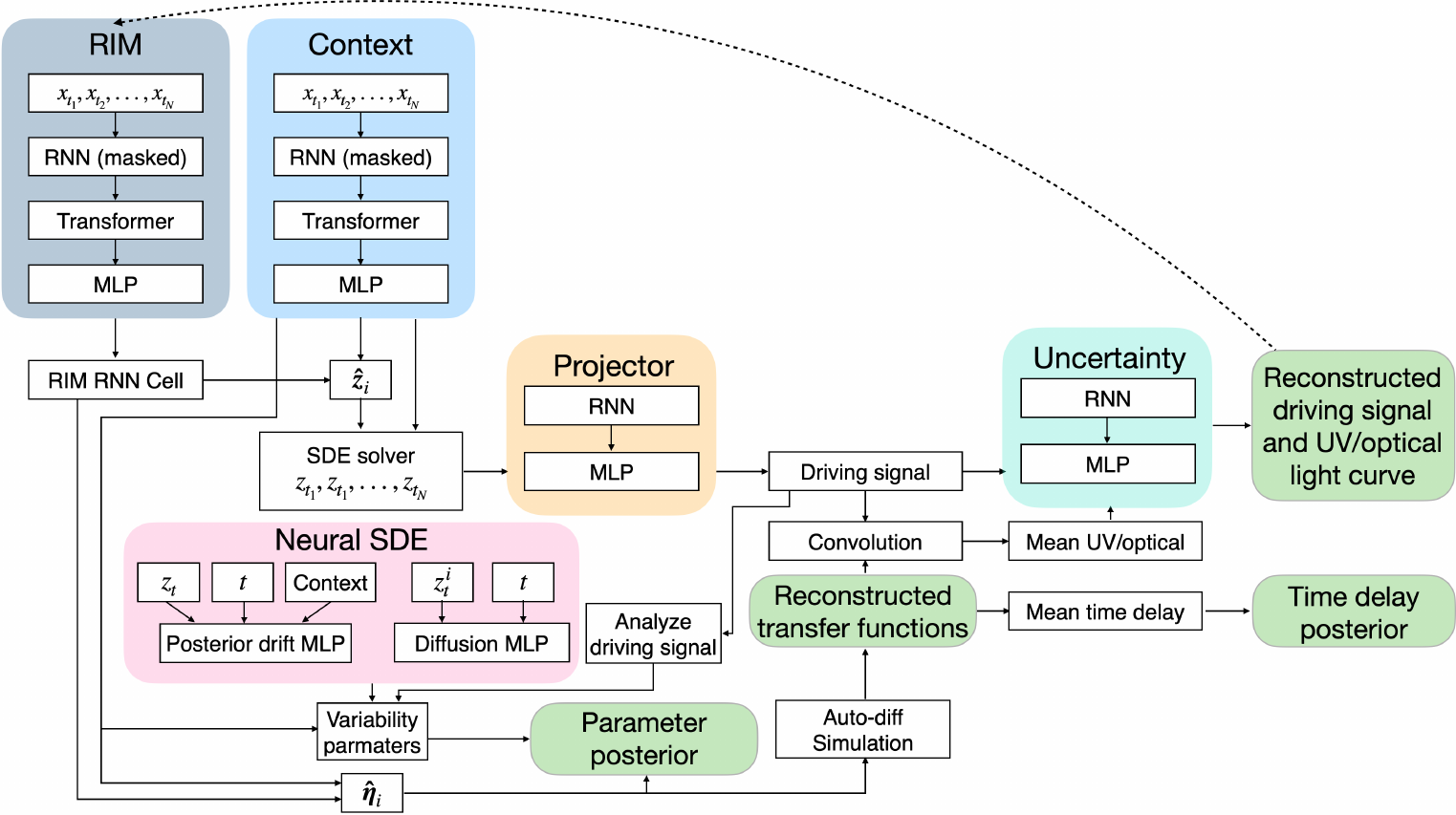}
    \caption{Diagram of our ML model. The data products are highlighted in green. The dashed line indicates the reconstructed UV/optical light curve that can optionally be used in the RIM procedure to update the predicted latent vector of the SDE $\boldsymbol{\hat{z}}_i$ and accretion disk parameters $\boldsymbol{\hat{\eta}}_i$. The convolution performed in the ML model is given by Equation~(\ref{eq:flux_reconstruction})}
    \label{fig:model_architecture}
\end{figure*}

A diagram of the model architecture is shown in Figure~\ref{fig:model_architecture}. The model has two main components to encode the light curve: the context network of the latent SDE, and the RIM network used only to adjust $\boldsymbol{\hat{\eta}}$ and $\boldsymbol{\hat{z}}$. Both contain bidirectional RNNs including GRU-D layer~\citep{GRUD}, a type of gated recurrent unit~\citep[GRU;][]{GRU} layer that is designed to handle the masking and irregular sampling. In addition, we combine the output of the RNN layers with a transformer encoder~\citep{vaswani2023attentionneed}. The GRU-D layers work well at handling the irregular sampling in the observed light curves by masking the unobserved time steps, while the transformers provide improved ability to handle long-term dependencies. Both the context and RIM networks require GRU-D layers to encode the irregularly sampled observations. The use of transformers also allows us to efficiently scale the model to many more parameters compared to RNNs. We use bidirectional RNN layers since they process the time series both forwards and backwards.

After the light curves are encoded, the context and latent vector $\boldsymbol{\hat{z}}$ are used by the neural SDE solver and then projected with an RNN to produce the reconstructed driving signal. The context and RIM are also used to predict the estimated accretion disk parameters $\boldsymbol{\hat{\eta}}$ that are used by the auto-differentiable simulation to produce the reconstructed transfer functions. The driving signal and transfer functions are then convolved within the ML model, numerically evaluating Equation~(\ref{eq:flux_reconstruction}) to produce the mean reconstructed UV/optical light curves. The mean reconstruction of the driving signal and UV/optical light curve is scaled, and the uncertainties are quantified using another RNN. In addition, the mean time delays are found from the reconstructed transfer functions, and the uncertainty is quantified. We also predict the variability parameters of the driving signal, using additional information from analyzing the driving signal such as fitting its PSD with five linear fits. The inferred driving and accretion disk parameters are combined, and the uncertainty is quantified. 

In summary, the ML model infers a posterior distribution on the accretion disk and variability parameters, a posterior of the time delay between wave bands, reconstructs transfer functions, and reconstructs the UV/optical light curve and the unobserved driving signal. Our model is built in \texttt{PyTorch}~\citep{Pytorch} and has 38,551,054 parameters, the majority of which come from the two transformers. More specific details about the NN architecture are given in Appendix~\ref{sec:appendix_architecture}.

\subsection{Uncertainty Qualification and Loss} \label{sec:loss}

We train our model with supervised learning by minimizing a weighted sum of four loss components. The driving variability is modeled as a latent SDE. Unlike in~\citet{Fagin_2024}, we do not sample the latent space of the SDE from a posterior distribution, but instead just directly predict $\boldsymbol{\hat{z}}$. We find this to significantly improve the performance of the light-curve reconstruction, since otherwise the reconstructed driving signal is not as adaptable. Our light-curve reconstruction is found by convolving the driving variability with the predicted transfer function kernels and scaling it to best match the observed data. The first component of the loss is the negative Gaussian log-likelihood between the reconstructed and true light curve, including the unobserved driving signal and averaged across the observed bands:
\begin{equation} \label{eq:NGLL}
\mathscr{L}_{\text{LC}} = \frac{1}{N} \sum_{t=1}^N \frac{(y_t-\hat{y}_t)^2}{2\hat{\sigma}_t^2} + \frac{1}{2}\log(2\pi\hat{\sigma}_t^2) \, , 
\end{equation}
where $y$ is the true light curve, $\hat{y}$ is the predicted mean, $\hat{\sigma}^2$ is the predicted variance, and $N$ is the number of time steps. We predict $\log(\hat{\sigma}^2)$ instead of $\hat{\sigma}$ for stability and to force the variance to be positive.

We include an additional component to the loss to ensure that our mean light-curve reconstruction closely matches the observations:
\begin{equation} \label{eq:context_loss}
\mathscr{L}_{\text{obs}} = \frac{1}{N_{\text{obs}}} \sum_{t=1}^N m_t \cdot \frac{(y_t-\hat{y}_t)^2}{2\sigma_{\text{LSST},t}^2} \, ,
\end{equation}
where $\sigma_{\text{LSST}}$ is the error of each observation, $m_t$ is a mask that is 1 if a band is observed at a time step $t$ and 0 if it is not, and \mbox{$N_{\text{obs}} = \sum_{t=1}^N m_t$} is the number of observations. The additional term in the negative log-likelihood $\frac{1}{2}\log(2\pi\sigma_{\text{LSST},t}^2)$ does not depend on our predictions and is therefore not included. 

For the parameter inference, we use a multivariate Gaussian mixture model. Each multivariate Gaussian has negative log-likelihood: 
\begin{equation} \label{eq:gaussian_log_likelihood}
\mathscr{L}_{\text{G},i} = \frac{1}{2} (\boldsymbol{y}-\boldsymbol{\hat{y}}_i)^\top \Sigma^{-1}_i (\boldsymbol{y}-\boldsymbol{\hat{y}_i}) + \frac{1}{2} \log(2\pi|\Sigma_i|) \, ,
\end{equation}
where $\boldsymbol{y}$ are the vectors of true values of our parameters, $\boldsymbol{\hat{y}}_i$ are the vectors of our predicted means, and $\Sigma_i$ are the covariance matrices of the prediction. To ensure that the covariance matrix is symmetric, positive semi-definite, and nonsingular, we predict the lower triangular matrices $L_i$ representing the Cholesky decomposition of the covariance matrices $\Sigma_i = L_iL_i^\top$. The diagonal of $L$ must be positive, so we take the softplus of each diagonal element. An $m\times m$ lower triangular matrix will have \mbox{$m(m-1)/2$} free parameters, with $m = 12$ for our case. We include into the loss the negative log-likelihood of the Gaussian mixture model:
\begin{equation} \label{eq:gaussian_mixture}
\mathscr{L}_{\text{param}} = -\log\left(\sum_{i=1}^{n}a_i e^{-\mathscr{L}_{\text{G},i} } \right) \, ,
\end{equation}
with mixture coefficients $a_i$ of our $n = 5$ multivariate Gaussians, normalized such that $\sum_{i=1}^n a_i = 1$. 

We additionally predict the relative time delay of each band with respect to the reference $i$ band. These time delays are based on our predicted transfer functions, but we also infer the lower triangular matrix to assign multidimensional uncertainty and include into the loss:
\begin{equation}
\mathscr{L}_{\tau} = \frac{1}{2}(\boldsymbol{\tau}-\boldsymbol{\hat{\tau}})^\top \Sigma^{-1}_{\tau} (\boldsymbol{\tau}-\boldsymbol{\hat{\tau}}) + \frac{1}{2} \log(2\pi|\Sigma_{\tau}|) \, ,
\end{equation}
evaluated even for unobserved bands.

The overall loss is the weighted sum of each term:
\begin{equation}
\mathscr{L}
= \mathscr{L}_{\rm LC} + \mathscr{L}_{\rm obs} + 2\,\mathscr{L}_{\rm param} + \mathscr{L}_{\tau} \, ,
\end{equation}
weighting $\mathscr{L}_{\rm param}$ twice to reflect our focus on parameter inference. When using a RIM, each term is additionally scaled by its iteration index to emphasize later iterations.

We use uniform priors for the variability and accretion disk parameters when simulating our light curves, given in Table~\ref{table:Parameter_table}. When training our ML model, we reparameterize the parameter labels from their physical values to between zero and one and then take the logit\footnote{$\text{logit}(x) = \log(\frac{x}{1+x})$ for $x \in [0,1]$} to scale them from $-\infty$ to $\infty$. We then evaluate the negative log-likelihood of our parameter posterior in this logit space. We take the sigmoid\footnote{$\text{sigmoid}(x) = \frac{1}{1+e^{-x}}$ for $x \in [-\infty,\infty]$} after drawing samples from the posterior, scaling the predictions back to between zero and one before scaling back to the original physical range. This transformation prevents any posterior probability from being wasted in physically impossible parameter space or outside the range of the training set (e.g., we must restrict the spin to \mbox{$-1 < a < 1$}). 

\begin{figure*}
    \centering
    \includegraphics[width=0.97\textwidth]{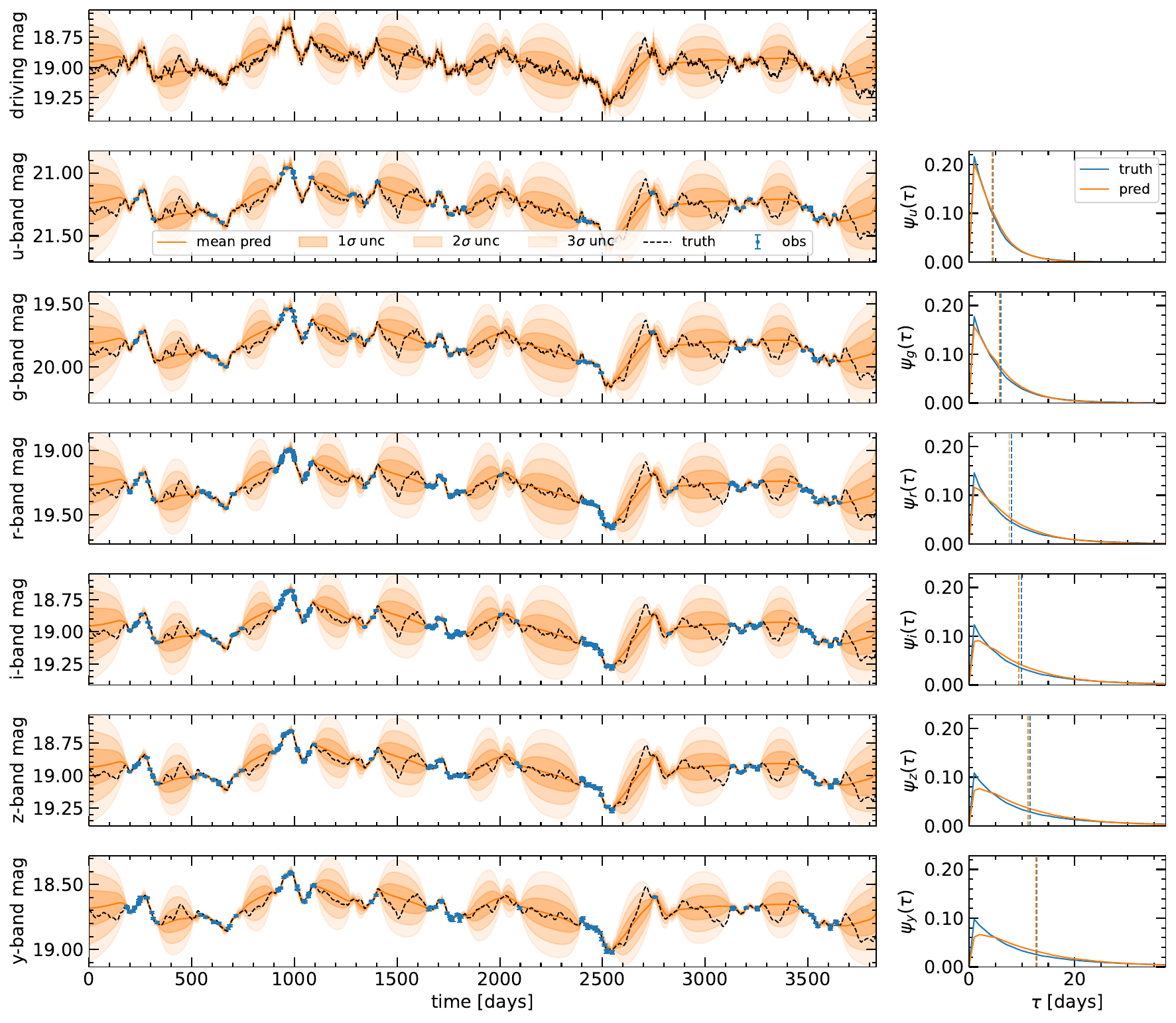}
    \caption{The left panels show an example reconstruction (orange) of a simulated 10.5 yr UV/optical quasar light curve (black) from our nominal test set with LSST-like observations (blue). The photometric errors are shown but are small for this bright quasar ($\sim0.01$ mag). The mean prediction is given by the orange solid line, with 1$\sigma$, 2$\sigma$, and 3$\sigma$ uncertainties highlighted. The UV/optical bands are reprocessed from the reconstructed driving signal (top panel) using the predicted transfer functions (right panels), with the mean time delays indicated by the dashed vertical lines.}
    \label{fig:example_reconstruction}
\end{figure*}

\subsection{Training} \label{sec:training}

We train our ML model with 100,000 light curves per epoch that are randomly regenerated on the fly. Regenerating the training set each epoch prevents overfitting by training with many more unique examples. This is especially important since our model has a large number of free parameters. We use fixed test sets of 10,000 light curves to evaluate the performance of our model after training.

Our ML model is trained for 30 epochs using the Adam optimizer~\citep{Adam}. We train with four A100 GPUs (80 Gb) in multiple stages. We first train without RIM to train faster. We use an initial learning rate of $8\times 10^{-4}$, determined through empirical tuning to balance convergence speed and stability. The learning rate is exponentially decayed by 0.95 each epoch, and a batch size of 26  per GPU (the maximum that fits in GPU memory). In the final four epochs, we use three RIM iterations with a batch size of 14 to test the RIM technique. Throughout training, we use gradient clipping with a maximum gradient norm of 250 to help prevent exploding gradients. Training took around 3 weeks. We did not employ any systematic hyperparameter tuning due to computational constraints in training the NN.

\section{Results} \label{sec:results}

\subsection{Light-curve Reconstruction Performance} \label{sec:light_curve_performance}

An example reconstructed UV/optical light curve, X-ray driving signal, and transfer function from the nominal test set is shown in Figure~\ref{fig:example_reconstruction}. Our ML model is able to properly quantify the uncertainty in its reconstruction, and it reconstructs the driving signal and transfer functions using just the UV/optical observations. Only the transfer functions from the mean parameter posterior that are convolved with the driving signal are shown. Although we do not directly infer uncertainty in the reconstructed transfer functions, an ensemble of reconstructed transfer functions could be produced by sampling from the parameter posterior.

\begin{table}
\begin{center}
\caption{Light-curve reconstruction performance of our latent SDE model compared to the GPR baseline in terms of the negative Gaussian log-likelihood given in Equation (\ref{eq:NGLL}) for six test sets with different types of driving signals. The values reported are the median $\pm$ median absolute deviation on the median across our test sets of 10,000 light curves. Lower is better and is shown in bold.}
\begin{tabular}{c c c}
Driving Signal & Latent SDE & GPR\\
\hline
BPL & $\boldsymbol{-1.528\pm 0.007}$ & $-1.409\pm 0.007$ \\
DRW & $\boldsymbol{-1.214\pm 0.007}$ & $-1.129\pm 0.007$ \\
BPL+sine & $\boldsymbol{-1.276\pm 0.007}$ & $-1.111\pm 0.007$ \\
Sine & $\boldsymbol{-1.847\pm 0.009}$ & $-1.765\pm 0.012$ \\
Sawtooth & $\boldsymbol{-1.221\pm 0.007}$ & $-0.971\pm 0.008$ \\
Square wave & $\boldsymbol{-1.098\pm 0.007}$ & $-0.933\pm 0.007$ \\
\end{tabular}
\label{table:latent_SDE_vs_GPR}
\end{center}
\end{table}

To evaluate the reconstruction performance of our model, we compare it to an exact multitask GPR baseline with DRW kernel on our test sets, the same baseline as in Section~4 of~\citet{Fagin_2024}. We use a GPR baseline since it is the standard method of reconstructing quasar variability~\citep[e.g.,][]{Stone_2022}, and has been previously used to compare performance to ML methods~\citep[e.g.,][]{Tachibana_2020,Danilov2022,Fagin_2024}.

To test the robustness of our ML model to irregular variability, we use additional test sets with different out-of-distribution driving signals. Table~\ref{table:latent_SDE_vs_GPR} reports the median $\pm$ median absolute deviation of the negative Gaussian log-likelihood across each test set of 10,000 light curves to provide an outlier-robust summary. We find that our ML model outperforms the GPR baseline for each driving signal, despite being trained only using the bended BPL. The exact parameter space of each driving signal is given in Appendix~\ref{sec:appendix_robustness}. Statistical significance is assessed by applying paired $t$-tests to the full set of 10,000 paired log-likelihood differences, yielding $p<10^{-6}$ for each test set. We choose to test our model on driving signals that are variable (BPL, DRW), quasi-periodic (BPL+sine), and periodic (sine, sawtooth, square wave). In each case, we are still able to predict the accretion disk parameters despite using out-of-distribution driving signals. We demonstrate this in Appendix~\ref{sec:appendix_robustness} and show that we are able to predict the black hole mass, redshift, temperature slope, and Eddington ratio similar to the case for the bended BPL. We also show an example light-curve reconstruction using a sawtooth driving signal to demonstrate how our model fits an out-of-distribution light curve.  

\begin{figure}
    \centering
    \includegraphics[width=0.47\textwidth]{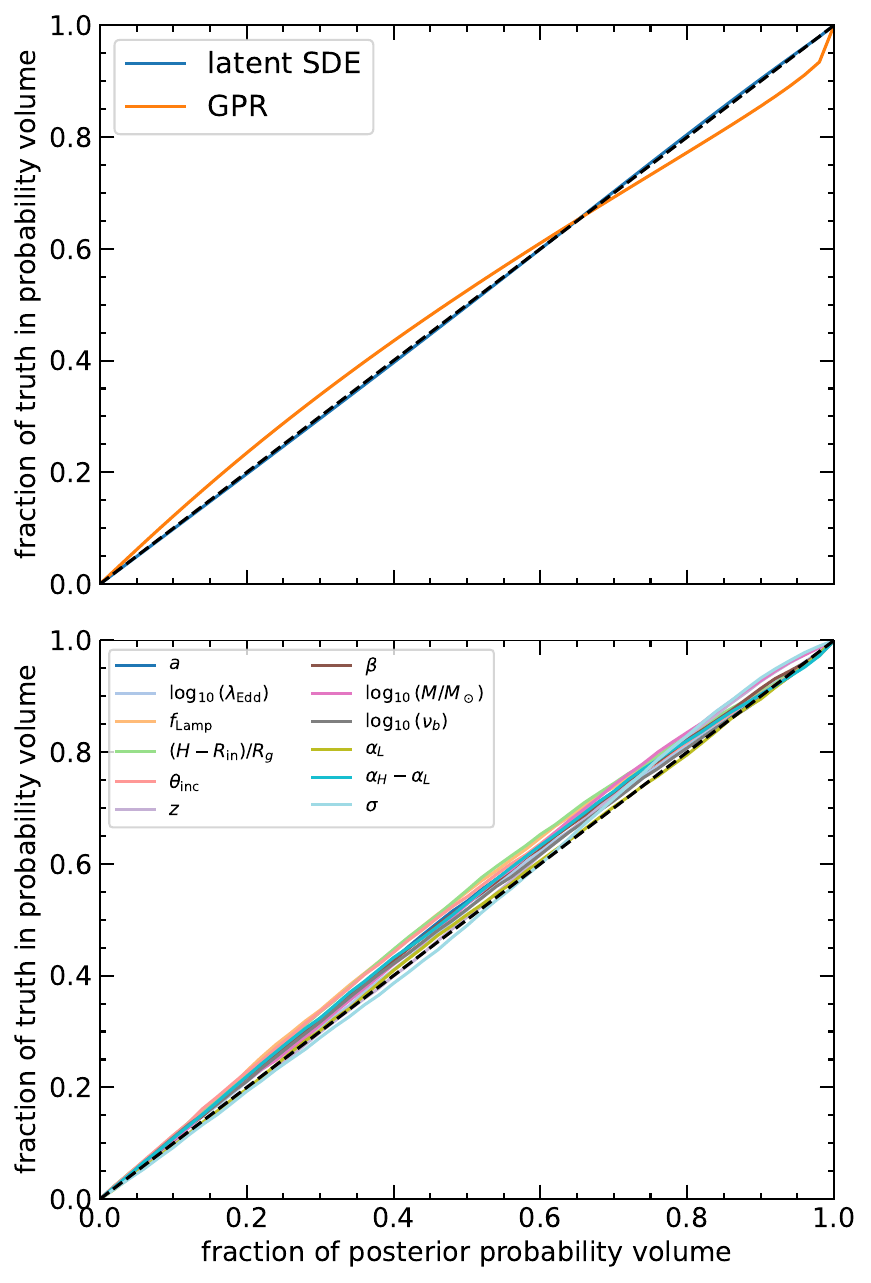}
    \caption{Evaluation of how well the uncertainties are calibrated for the light-curve reconstruction (top panel) and parameter inference (bottom panel) by showing the fraction of the truth encompassed within the posterior probability volume across the nominal test set. Perfect uncertainty calibration is shown by the black dashed line along the diagonal.}
    \label{fig:coverage_prob}
\end{figure}

In the top panel of Figure~\ref{fig:coverage_prob}, we show that the uncertainties we predict in our UV/optical light-curve reconstruction are well calibrated for the nominal test set, while the GPR baseline is misaligned. We may expect the GPR baseline to be misaligned, since it assumes a DRW kernel, while our driving signal is generated from a more general BPL PSD and convolved with the transfer function kernels. 

\subsection{Parameter Inference Performance} \label{sec:parameter_performance}

\begin{figure*}
    \centering
    \includegraphics[width=0.97\textwidth]{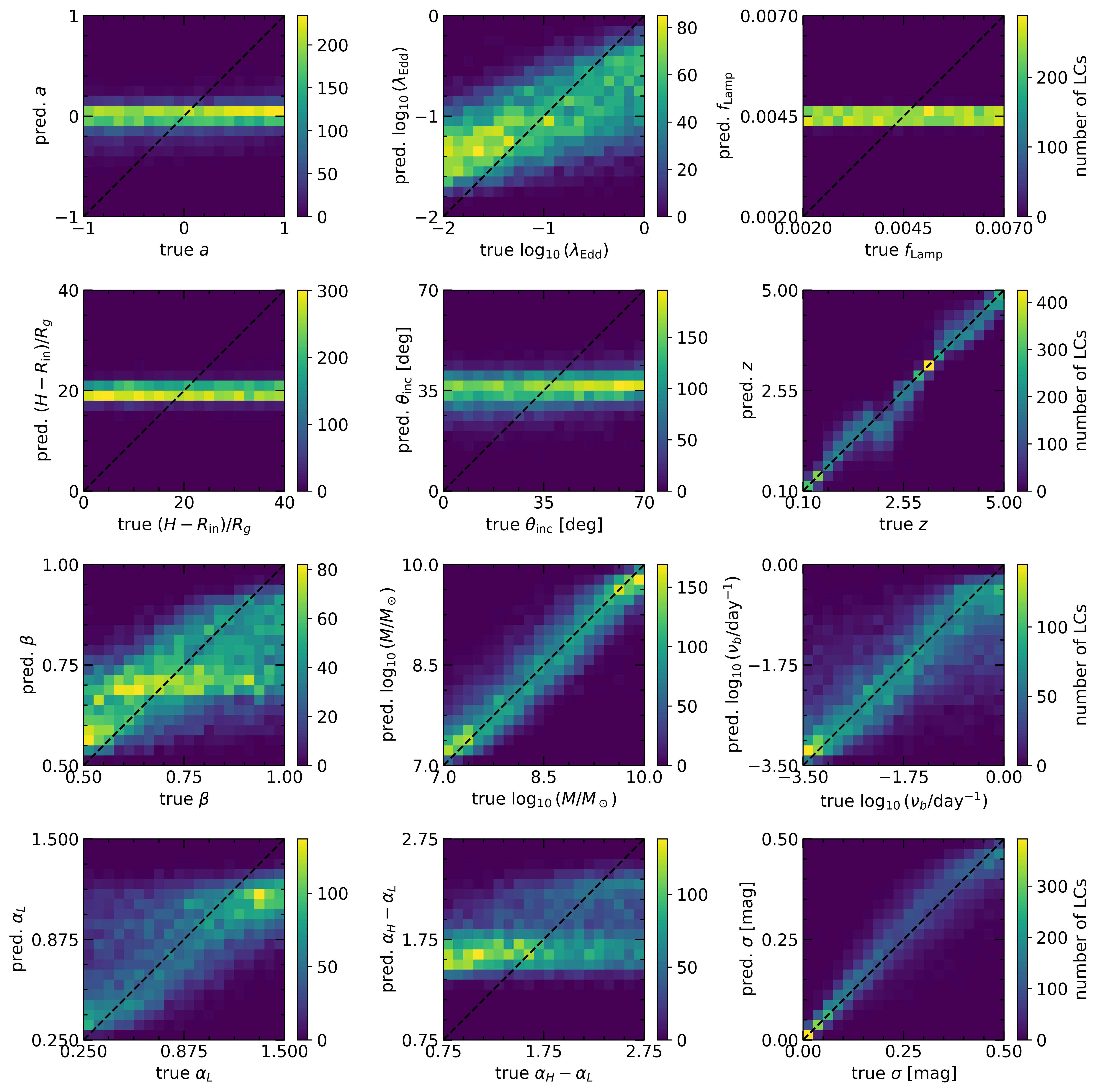}
    \caption{Median prediction compared to the true value for each predicted parameter (given in Table~\ref{table:Parameter_table}) across the nominal test set. The ideal case where the median prediction matches the truth is given by the black dashed line across the diagonal.}
    \label{fig:median_pred_vs_truth}
\end{figure*}

In Figure~\ref{fig:median_pred_vs_truth}, we show the median prediction compared to the true value for each parameter across the test set, demonstrating the ability of our model to predict each parameter. In this figure, we only show the median, but a full multivariate posterior is predicted for each light curve (see, e.g., Figure~12 of~\citet{Fagin_2024}). In the bottom panel of Figure~\ref{fig:coverage_prob}, we show that the uncertainties we predict in our parameter posteriors are well calibrated, although overall slightly underconfident.

For the accretion disk parameters, we can predict the mass and redshift with high accuracy and place meaningful constraints on the temperature slope and Eddington ratio. The ML model struggles to predict the inclination angle and black hole spin but can make some constraints, while the corona height and lamppost strength cannot be constrained at all. We show the influence that each parameter has on the transfer functions in Appendix~\ref{sec:appendix_parameters}, which explains why some of the accretion disk and black hole parameters can be predicted more easily than others, although there is additional information in the mean brightnesses. The mass and temperature slope predictions are better than in~\citet{Fagin_2024} due to modeling the mean brightness of each band, and we can now also predict the redshift and constrain the Eddington ratio. Our ML model cannot predict the inclination angle well because the inclination angle has almost no effect on the mean time delays, except at very high inclinations. While there is a higher-order effect of the inclination on the standard deviation of the transfer function, the associated brightness suppression is less informative, as we properly model the brightness of each band. The difficulty in constraining the inclination angle is primarily why the light curve and mean time delays can be well reconstructed while the exact shape of the transfer functions can still be off, such as in Figure~\ref{fig:example_reconstruction}. The shape of the predicted redshift is not continuous due to the $u$ and $g$ bands being suppressed at high redshift (starting at $z = 2.29$), causing the predictions to be very accurate at specific redshifts where these bands are being cut off. In addition, we are able to predict the bended BPL PSD parameters of the driving signal, although it is more challenging than the DRW parameters~\citep[see][]{Fagin_2024}. This is due to the bended BPL law having two additional parameters corresponding to the low- and high-frequency limits of the power spectrum, and the parameters are approximately degenerate. We note that we restrict the posterior distribution to be within the uniform prior, as mentioned in Section~\ref{sec:loss}. In Figure~\ref{fig:median_pred_vs_truth} we only show the median predictions, which by design will never reach the edges of the parameter space. In the worst case, the ML model will predict a median at the center of the prior, as is the case for the corona height and lamppost strength. 

We demonstrate in Appendix~\ref{sec:appendix_robustness} that our model can still infer the black hole mass with out-of-distribution driving signals, and the performance of all accretion disk parameters remains consistent with the results shown in Figure~\ref{fig:median_pred_vs_truth}. In addition, we quantify the root mean squared error (RMSE) of each parameter for each test case. For instance, the nominal RMSE is 0.33 for $\log_{10}(M/M_\odot)$ and 0.32 for the redshift.

\subsection{Example Relative Time Delay Inference} 

\begin{figure*}
    \centering
    \includegraphics[width=0.85\textwidth]{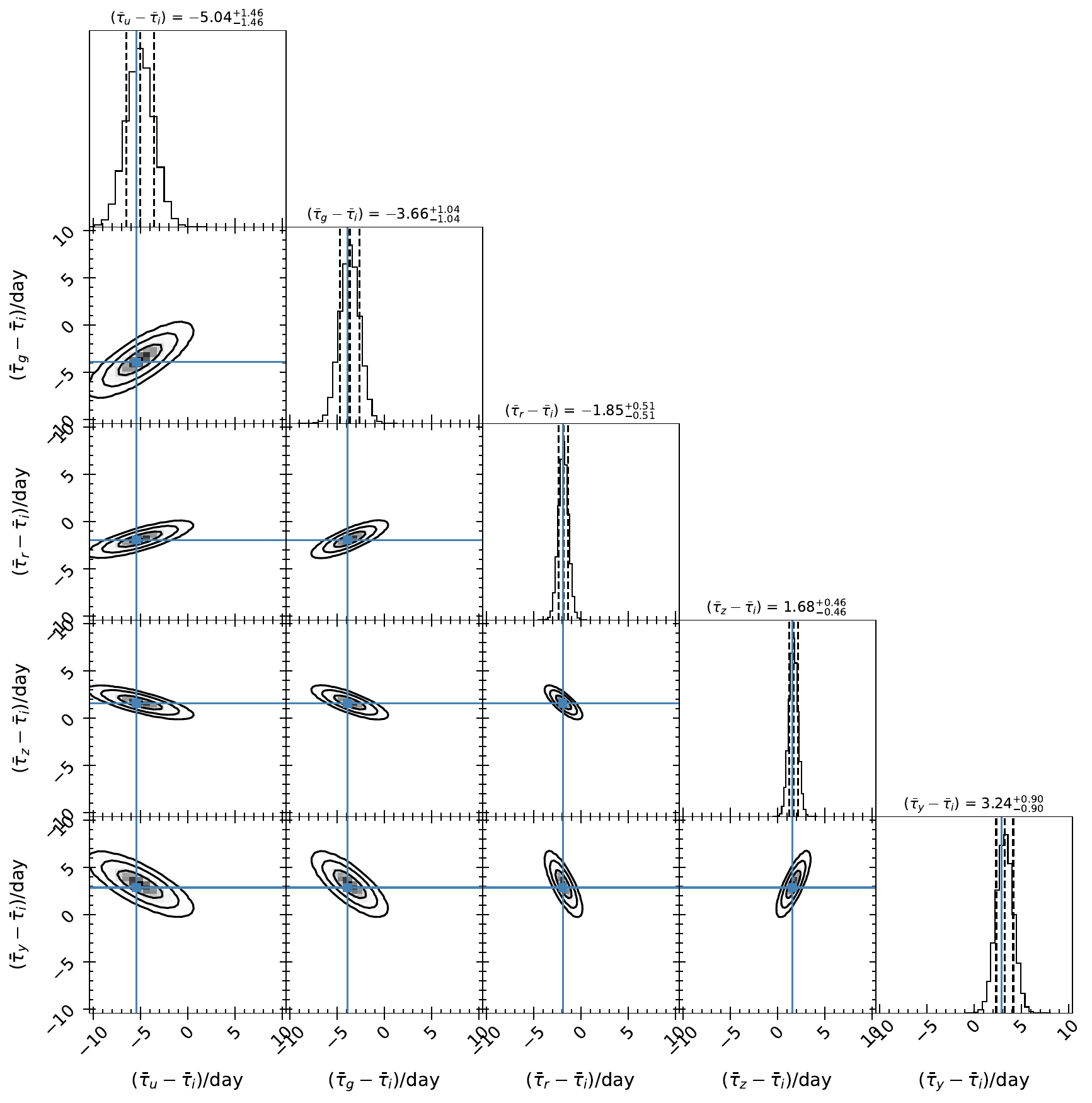}
    \caption{Example predicted posterior distribution for the time-delay measurements using a reference $i$ band for the same test light curve shown in Figure~\ref{fig:example_reconstruction}. The mean predictions come from the reconstructed transfer functions, while the uncertainties are separately quantified by our ML model. The diagonal elements display marginal distributions, with the median and 1$\sigma$ levels indicated by dashed lines. The central elements depict 1$\sigma$, 2$\sigma$, and 3$\sigma$ contour levels. The true time delays are overlaid in blue.}
    \label{fig:time_dealy}
\end{figure*}

The mean time delays between bands are found directly from the mean times of our predicted transfer functions so that our model is entirely self-consistent. The time-delay differences are defined with respect to the $i$ band as a reference. Our model predicts time delays even in the case where we do not observe the bluest bands, since we still have the transfer functions corresponding to our predicted accretion disk parameters. Our ML model then produces uncertainties associated with the mean time delays by predicting the lower triangular matrix $L$ of its covariance matrix $\Sigma = LL^\top$. An example posterior of the time-delay measurements is shown in Figure~\ref{fig:time_dealy}. The time-delay posterior is directly inferred by our ML model, instead of requiring MCMC sampling.

\subsection{Recurrent Inference Machine Evaluation} 

We evaluate the performance of our model using RIM for three iterations. The average loss across the nominal test set does decrease each iteration, given by: $0.619$, $0.580$, $0.579$. Therefore, the RIM procedure has a minor improvement in the performance of the model compared to using just a single iteration.

\section{Discussion} \label{sec:discussion}

\subsection{Comparison to Traditional MCMC-based Methods}

We constructed the first ML approach to model the UV/optical light curve, driving variability, accretion disk reprocessing transfer functions, and time delays in a single unified framework by embedding the accretion disk reprocessing model into our NN architecture. This makes our predictions more interoperable compared to modeling each process individually. Our method enables the fast inference of accretion disk and variability parameters, time delays between wave bands, and the reconstruction of the driving signal and UV/optical bands.

Unlike traditional methods of measuring time delays, our model can use information related to the mean brightness and variability to inform our ML model on the accretion disk parameters and time-delay measurements. We also model the driving variability in a free-form way (i.e., using a latent SDE) instead of using a DRW process like \texttt{JAVELIN} or a Fourier series like \texttt{CREAM}. In addition, we directly predict the variability parameters of the driving signal without requiring it to be a Gaussian process. Furthermore, we can predict more parameters of the accretion disk, including the mass, Eddington ratio, redshift, temperature slope, and variability parameters of the unobserved driving signal. Comparatively, \texttt{CREAM} typically only measures $M\dot{M}$. 

While our new method is more computationally demanding to train than~\citet{Fagin_2024} due to having many more parameters and the use of a RIM, during inference it is still quick enough to easily apply to the tens of millions of quasar light curves expected from LSST in several hours. Specifically, the average inference time on our GPUs using a batch size of 64 is about 27 minutes per million light curves using three iterations of the RIM, or 7 minutes using one iteration. Comparatively, using \texttt{JAVELIN} or \texttt{CREAM} on tens of millions of light curves would be infeasible. This is currently a major limitation of traditional MCMC-based methods; however, faster methods may be developed that take advantage of GPU acceleration to overcome this limitation.

Instead of being model dependent, we could also use analytic functions such as using simple top-hat transfer functions like \texttt{JAVELIN} defined with respect to the bluest band (i.e., only five transfer functions for six bands since the bluest band is treated as the effective driving variability). One may also reconstruct the effective transfer functions in a free-form manner, but this approach would require regularization or the use of a complete set of basis functions to make the inverse problem tractable. In such cases, the X-ray driving variability would not be reconstructed, which was a major goal of this work, but these methods could be explored in future studies. 

\subsection{Limitations of This Work}

In this work, we did not account for time delays that may arise from the broad-line or narrow-line regions of quasars. Modeling the BLR is difficult since its geometry is not well known. Including the BLR into our simulation would introduce a large number of additional parameters~\citep{best2024amoebaagnmodeloptical}. The time delays caused by the BLR can cause our method to overestimate the recovered black hole mass, since the BLR can introduce larger time lags. However, this is also the case for traditional sampling methods like \texttt{JAVELIN} and \texttt{CREAM}. Furthermore, there could be a radially dependent albedo or color-correction factors to the blackbody radiation. In addition, the thin-disk NT model may break down at high Eddington ratios, and different disk models such as the slim-disk model~\citep{SlimDisk} may be more accurate. Some of the deviations from the thin-disk model may already be partially accounted for by the inclusion of the accretion rate wind model (Equation~(\ref{eq:wind})) that modifies the slope of the temperature profile. There may also be long negative time lags at the viscous timescale of the disk~\citep{Yao_2023,secunda2024negativelagsviscoustimescale}, which our model could be adapted to predict and could be used to study the vertical structure of the disk. We also assumed that each light curve is coming from type 1 quasars, but there may be contamination from type 2 quasars that are misidentified. We tested that our model is robust to different out-of-distribution driving signals, but it would be useful to test how our pretrained model behaves for different disk models. Moreover, we assumed that each driving signal was generated from a stationary process, but there could be nonstationary variability such as from flaring or tidal disruption events.

In our training set, we assumed we had the full 10 yr of LSST data. In future work, we plan to evaluate the model's performance using different stages of the survey, such as using only the first 1, 3, and 5 yr of LSST data. We may also compare the performance of our model between light curves from the Wide Fast Deep and Deep Drilling Fields. Furthermore, a fraction of the LSST quasar light curves will have observations from previous surveys that can be combined with LSST data to lengthen the light curve.

We find the driving signal parameters of the BPL to be more difficult to predict than the DRW parameters used in~\citet{Fagin_2024}. This is because there are two additional parameters, and the additional degrees of freedom introduce degeneracies. Here we parameterized the driving variability as a bended BPL, since it has been found to fit X-ray variability data. Some authors such as~\citet{papoutsis2024xrayreverberationexplanationuvoptical} fix the lower-frequency slope to $\alpha_L = 1$, which would get rid of degeneracies, and our predictions for $\nu_b$ and $\alpha_H$ would improve. We choose to keep the bended BPL more general to account for the wide range of variability possible in LSST. We could, however, use any parameterization of the PSD to train our model. 

\subsection{Discussion of Machine Learning Architecture}

As far as we are aware, this work represents the first application of transformers in modeling quasar variability. To process the irregularly sampled time series, we first use GRU-D and GRU layers and then transformer encoders concatenated with the output. The GRU-D handles the irregular sampling more effectively than transformers, which we found struggled to converge on their own. By combining both RNN and transformer architectures, we achieved stable convergence while leveraging the transformer's improved handling of long-term dependencies in lengthy time series. This approach also enabled us to scale our model to tens of millions of parameters, making it the largest deep learning model applied to quasar variability to date. 

In future work, we may test improvements to the NN architecture. One major modification could be to reconstruct the light curve by repeatedly sampling driving signals and accretion disk parameters from the latent space. This would work like a variational auto-encoder to yield nonparametric posterior distributions. However, with our current training paradigm this would require repeated sampling during training to obtain approximate posterior distributions to use in the loss function, and this would be too computationally demanding to train. It could also be possible to replace all RNNs with just a single transformer encoder. Furthermore, we could explore other encoder architectures~\citep[e.g.,][]{schirmer2022modeling}. In this work, we bin the time series and transfer functions to daily time intervals. Ideally, we would use smaller bins, but we are limited by GPU memory and training time. Using daily intervals could negatively impact the parameter inference of quasars with smaller-mass black holes ($M \lesssim 10^{7}M_\odot$), where the time delays can be on the order of 1 day or less.

We use our framework to iteratively improve the light-curve reconstruction and accretion disk parameter estimation with a RIM by analyzing the residuals between the predicted light curve and observations. We found the RIM to have only a minor advantage compared to using a single iteration. This is likely due to several factors. For example, we pretrained our model using a single iteration to save on training time. We also only use the RIM to adjust the mean accretion disk parameters and latent vector of the SDE, but perhaps all parameters should be iteratively adjusted.  In future work, we plan to explore using a single encoder, instead of separate context and RIM modules, to iteratively adjust the entire context. Here we use a very large model with tens of millions of parameters, and it may be the case that a minimum is achieved with only a single iteration. In addition, our time series are stochastic, irregularly sampled, and noisy. Therefore, we cannot reconstruct the time series to the pixel level, as is the case for image reconstruction tasks where a RIM has traditionally been applied. The RIM technique may be more advantageous for well-sampled light curves, such as in the Deep Drilling Fields.

\subsection{Outlook and Future Applications}

Once LSST data becomes available, we can fine-tune our model using the data with self-supervised learning. This would help generalize our ML model trained with simulated data and supervised learning to the real LSST observations. This could be done simply by minimizing the loss with respect to how well our reconstructions match LSST observations (similar to Equation~(\ref{eq:context_loss})). The light-curve reconstruction and parameter inference are linked through our auto-differentiable simulation of the accretion disk, so by fine-tuning the model on the observations, we can potentially improve all aspects of our predictions. Furthermore, semi-supervised learning could be used if there is a partial set of LSST light curves that have some parameter estimates from spectra. In addition, the parameters of the BPL driving signal may be related to properties of the accretion disk and black hole~\citep{arevalo2023universalpowerspectrumquasars}. In our training set, we choose to keep them independent so that these correlations could potentially be measured by our model without bias. However, our model could improve its predictions by finding these correlations when fine-tuning with the real data. Fine-tuning our model with data from the Deep Drilling Field quasar sample would be particularly beneficial. In future work, we could also explore adapting our method for fully self-supervised training.

The auto-differentiable simulation we develop enables new physics-based ML methods like ours to be developed to model quasar variability. In addition, more traditional methods that rely on gradient-based sampling, such as Hamiltonian MCMC~\citep{betancourt2018conceptualintroductionhamiltonianmonte}, can be developed using our simulation. Hamiltonian MCMC can greatly speed up the rate of convergence compared to traditional MCMC by using gradient-based guidance. Even without the use of auto-differentiation, our simulation can take advantage of GPU acceleration to speed up traditional MCMC methods. We benchmark the speed of our simulation with and without GPU acceleration. For a single transfer function, we find the GPU to be $1.5\times$ faster than on the CPU (taking 0.54 s compared to 0.81 s). With batches of 50, we find this improves to a $67\times$ speedup on a single GPU (taking 0.55 s per batch compared to 36.9 s). The batch size can be scaled to as many samples that can fit in GPU memory for maximum speedup. The driving variability can also be generated using \texttt{PyTorch}, and is included in the codebase although not directly used in this work. 

Our accretion disk simulation includes a wind-based accretion rate and GR effects, and it can be readily adapted and expanded upon in future work. Our framework enables hypothesis testing by training multiple ML models, each incorporating a different disk model, and evaluating their inferred parameters and time delays against both observational data and GR magnetohydrodynamics (GRMHD) simulations. In observational data, these quantities can be measured through independent methods, providing a complementary means of assessing model accuracy. In GRMHD simulations, the true underlying values are known, allowing for a direct comparison between model predictions and physically motivated disk structures~\citep[e.g.,][]{Koudmani_2024}. This approach provides a systematic way to evaluate competing accretion disk models and refine our understanding of disk variability. Our framework supports the integration of any auto-differentiable disk models, ensuring flexibility in exploring a wide range of accretion physics scenarios. Even complex simulations could be made auto-differentiable by training a NN to emulate them and embedding the fixed-weight pretrained model into our NN architecture. In addition, we model the quasar spectrum and integrate the mean brightness across the LSST response functions to obtain physically consistent mean brightnesses of each wave band. This is a major upgrade from~\citet{Fagin_2024}, since this information can be used to break parameter degeneracies and infer photometric redshifts. This is especially important in cases with low variability.

The framework developed in this work offers significant potential for anomaly detection in various contexts. For instance, the latent space of the SDE can be used in an unsupervised way to identify out-of-distribution sources of variability by comparing it to the typical distribution of quasar light curves or employing methods like isolation forests. Additionally, by analyzing the reconstructed driving signal, we may detect quasi-periodic signals that could indicate the presence of a supermassive black hole binary. The model could also be extended through supervised learning to classify light curves as belonging to phenomena such as supermassive black hole binaries, changing-look AGN, flaring events, tidal disruption events, or gravitational microlensing. 

LSST will observe tens of millions of quasars, and only a small fraction of them will get follow-up spectra to obtain precise redshift measurements. Therefore, most quasars will rely on photometric redshifts. In theory, the photometric redshift predictions of our ML model could outperform simple photometric redshift estimators based on the brightness differences between wave bands. This is because the photometric redshifts should depend most strongly on the asymptotic brightnesses of each band, and our model incorporates modeling the time-variable brightness. Furthermore, the redshift affects the time delays and driving variability time scale, so by analyzing all these processes in a unified framework, we can outperform methods relying on static data. Quasars that do have follow-up spectra can be better constrained by using the redshift as a prior in our ML model. The mass and other properties of the accretion disk may be constrained from measurements of the broad-line region~\citep{Panda_2019}, which could also be used to inform our ML model. 

Some of the accretion disk parameters are difficult to predict for individual quasars. Hierarchical inference could be used with the entire LSST sample to estimate the population-level distributions of the parameter space~\citep{Wagner_Carena_2021}. For example, the population-level distribution of the temperature slope $\beta$ could be used to test accretion disk and wind models, despite being challenging to constrain for individual quasars with only six bands. Furthermore, we could measure the relationship between driving variability parameters $\nu_b$, $\sigma$, $\alpha_L$, and $\alpha_H$ and black hole properties like the mass.

\section{Conclusions} \label{sec:conclusion}

Our model fits the UV/optical variability, reconstructs the driving variability signal, predicts accretion disk and variability parameters, and measures the relative time delay between bands, all self-consistently using a ML model that incorporates the physics of the accretion disk reprocessing into its architecture using an auto-differentiable simulation and latent SDEs. We incorporate transformers into our model to enhance its capacity for capturing long-term dependencies and to scale it to tens of millions of parameters. Our method is ready to be applied to the entire sample of tens of millions of monitored LSST quasars in a matter of hours. In comparison, using GPR frameworks such as \texttt{Celerite}~\citep{celerite} and traditional curve-shifting techniques like \texttt{JAVELIN}~\citep{Zu_2011,Zu_2016} and \texttt{CREAM}~\citep{Starkey_2015} will be computationally infeasible. Furthermore, we test the robustness of our ML model by comparing the performance of our pretrained model on light curves with out-of-distribution driving signals including variable, quasi-periodic, and periodic signals. We find our model outperforms a multitask GPR baseline in all cases and can still infer the accretion disk parameters. 

We aim to incorporate as much physics into our ML model as possible. The latent SDE captures the physics of the stochastic driving variability, while the auto-differentiable simulation models the reprocessing of the driving signal on the accretion disk. In previous works, there has been no mechanism to ensure that the inferred accretion disk parameters correspond to the time delays in the light-curve reconstructions~\citep{JiWon_2021,Fagin_2024}. In addition, our model fits the power spectrum of the reconstructed driving signal with several linear segments to better estimate the driving signal parameters. By embedding these physical processes into the NN, we achieve a model that is more robust and interpretable compared to traditional black-box parameter estimators. Furthermore, by linking these processes the model can refine its parameter estimates when using self-supervised learning on real observations, such as those from LSST. On real data, we can also compare the predicted time delays of our model to those obtained using traditional curve-shifting techniques, providing a way to validate the time-delay measurements and search for anomalies. Additionally, this method can test reprocessing models by comparing the reconstructed driving signals to X-ray data. The ML approach we present is highly general and can be adapted to other multivariate time series with irregular sampling, particularly for blind deconvolution or inverse problems. Our ML model and auto-differentiable and GPU-accelerated accretion disk simulation are open-sourced and available on GitHub\footnote{\href{https://github.com/JFagin/Quasar_ML}{https://github.com/JFagin/Quasar\_ML}}.

\section*{Acknowledgments}

Support was provided by Schmidt Sciences, LLC. for J.F., J.C., H.B., and M.O. H.B. acknowledges the GAČR Junior Star grant No. GM24-10599M for support. The authors would like to thank Sajesh Singh for computing support and Matthew Temple for useful discussions. We thank the anonymous referee for useful comments. This work used resources available through the National Research Platform (NRP) at the University of California, San Diego. NRP has been developed, and is supported in part, by funding from National Science Foundation, from awards 1730158, 1540112, 1541349, 1826967, 2112167, 2100237, and 2120019, as well as additional funding from community partners. M.J.G. acknowledges support from National Science Foundation award AST-2108402. K.E.S.F. acknowledges support from NSF AST-2206096, NSF AST-1831415, and Simons Foundation grant No. 533845.

\software{
\texttt{PyTorch}~\citep{Pytorch},
\texttt{torchsde}~\citep{Torch_SDE}, 
\texttt{BoTorch}~\citep{BoTorch},
\texttt{Matplotlib}~\citep{Matplotlib}, \texttt{Numpy}~\citep{Numpy}, 
\texttt{SciPy}~\citep{Scipy}, 
\texttt{Astropy}~\citep{Astropy}, 
\texttt{corner.py}~\citep{corner}, 
\texttt{speclite}~\citep{speclite}.
}

\appendix

\section{Novikov-Thorne Model} \label{sec:appendix_NT_model}

The disk dimensionless flux factor in the SS thin-disk model~\citep{Shakura_1973} is:
\begin{equation}
f_{\text{SS}}(R,a) = \frac{R_g^3}{R^3}\left(1-\sqrt{\frac{R}{R_{\text{in}}}}\right) \, ,
\end{equation}
with radius on the disk $R$, gravitational radius $R_g = GM/c^2$, inner radius $R_{\text{in}}$, dimensionless spin of the black hole  $a$, and flux $F_{\text{SS}} = (3GM\dot{M}/8\pi R_g^3) f_{\text{SS}}(R, a)$.
The NT model~\citep{Novikov_1973} includes GR corrections, and its dimensionless flux factor can be expressed analytically as:
\begin{multline}
f_{\text{NT}}(R, a) = \left(x^7-3x^5+2ax^4\right)^{-1} \left[x-x_0-\frac{3}{2}a\log\left(\frac{x}{x_0}\right)
-\frac{3(x_1-a)}{x_1(x_1-x_2)(x_1-x_3)}\log\left(\frac{x-x_1}{x_0-x_1}\right) \right. \\ \left.
-\frac{3(x_2-a)}{x_2(x_2-x_1)(x_2-x_3)}\log\left(\frac{x-x_2}{x_0-x_2}\right)-\frac{3(x_3-a)}{x_3(x_3-x_1)(x_3-x_2)}\log\left(\frac{x-x_3}{x_0-x_3}\right) \right] \, ,
\end{multline}
for $x = \sqrt{R/R_g}$, $x_0 = \sqrt{R/R_{\text{in}}}$, $x_1 = 2\cos\left(\frac{1}{3}\cos^{-1}(a)-\frac{\pi}{3}\right)$, $x_2 = 2\cos\left(\frac{1}{3}\cos^{-1}(a)+\frac{\pi}{3}\right)$, and \mbox{$x_1 = -2\cos\left(\frac{1}{3}\cos^{-1}(a)\right)$} with flux $F_{\text{NT}} = (3GM\dot{M}/(8\pi R_g^3) )f_{\text{NT}}(R, a)$. Both models are very similar and the choice between the two has only a small effect on the transfer functions.

\section{Impact of parameters on transfer functions} \label{sec:appendix_parameters}

\begin{figure}
    \centering
    \includegraphics[width=0.97\linewidth]{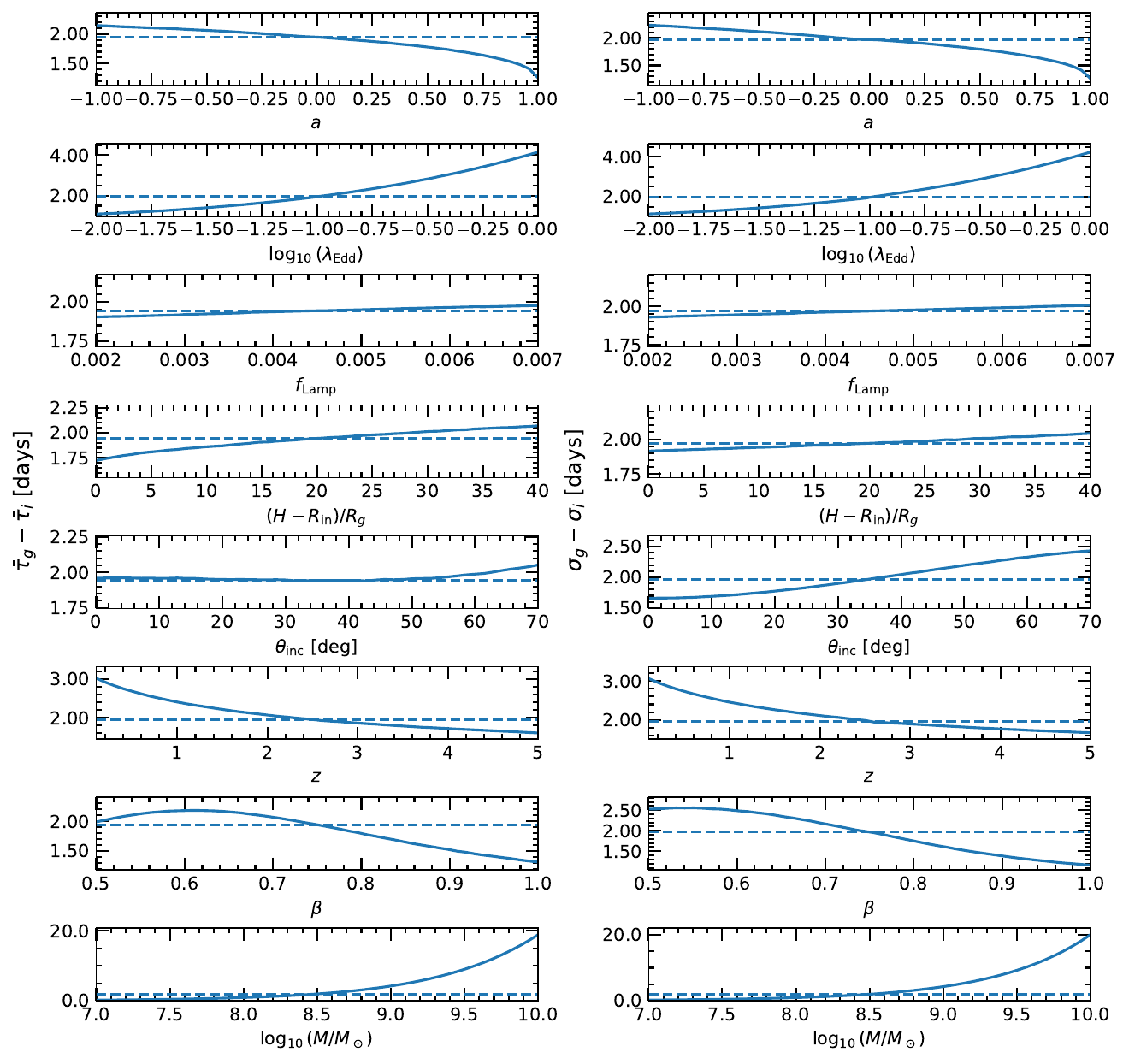}
    \caption{Time delays (left panels) and standard deviation (right panels) between the $g$ and $i$ bands as a function of each parameter in our accretion disk simulation while the other parameters remain fixed to the mean values of the prior (see uniform priors in Table~\ref{table:Parameter_table}). The horizontal dashed line indicates the time delay and standard deviation when all the parameters are fixed to the mean values of the prior.}
    \label{fig:time_delay_parameters}
\end{figure}

Figure~\ref{fig:time_delay_parameters} presents the mean time delays and standard deviation differences between the $g$ and $i$ bands of the transfer functions as each parameter is varied. The mean time delay reflects the relative shift between the two bands, while the standard deviation affects the relative smoothness of the variability. We focus on the differences between bands, since only variations in time delay or smoothness are observable. This figure highlights the influence of each parameter on the observed variability; however, changes in the mean brightness of each band coming from modeling the spectrum help to break some parameter degeneracies. It is apparent that the black hole mass has the largest effect, followed by the Eddington ratio, redshift, and temperature slope. This is consistent with our ML model's ability to predict these parameters, as demonstrated by Figure~\ref{fig:median_pred_vs_truth}.

\section{Machine learning architecture} \label{sec:appendix_architecture}

Here we expand on the description of the NN architecture from Section~\ref{sec:model_architecture}. The encoders of the NN contain bidirectional RNNs and begin with an initial GRU-D layer~\citep{GRUD} followed by two GRU layers~\citep{GRU}. Each layer is made bidirectional by splitting each RNN layer into two RNNs with half the hidden size each: one that processes the time series forwards and the other backwards. The output of the backwards RNN is flipped along its time axis and concatenated with the output of the forward RNN. We also use a transformer encoder after the RNN layers that is concatenated with the output~\citep{vaswani2023attentionneed}. The output of the RNNs and transformers is followed by two fully connected layers. 

The input of the context network is the observed light curve, padded for an additional 800 days to account for the extra time needed in the reconstructed driving signal due to the length of the transfer function kernels lost when performing the convolution (see Equation~(\ref{eq:flux_reconstruction})). For the RIM network, we start with a constant predicted light curve at the mean of each band. The input to the RIM network is the predicted light curve $\hat{y}$, the observations $y$ with variance $\sigma^2$, the gradient of the observations $-2\bar{\sigma}^2\cdot m\cdot (y-\hat{y})/\sigma^2$, the $z$-score $\bar{\sigma} \cdot m \cdot (y-\hat{y})/\sigma$, and the log-likelihood $\bar{\sigma} \cdot m \cdot (y-\hat{y})^2/2\sigma^2$, where $m$ masks all the unobserved points and $\bar{\sigma}$ is the root mean variance of the observational noise, which is included for stability so each input has units of magnitude. For the context network, we use the entire time vector for the latent SDE. We also have a two-layer fully connected layer to aggregate across the time dimension of the context and RIM networks by taking as input the first and last times and the mean and standard deviation across the time dimension. The output of the RIM network goes into a small network that is one GRU cell and a fully connected layer, with the hidden state of the GRU cell being updated with each RIM iteration. The context is also used to predict the parameters, the latent space of the SDE, and the normalization of the light curve. 

There is an additional RNN without the transformer to produce uncertainty in the reconstructed driving and UV/optical variability with the same architecture but no transformer. We use one more RNN to project the output of the latent SDE to the mean and standard deviation on the observation space~\citep[see][]{Fagin_2024}. This RNN has a linear skip connection between the output of the SDE and a two-layer bidirectional GRU-based RNN, so the RNN layers can be easily bypassed if they are unnecessary.

Multi-layer perceptrons (MLPs) produce the posterior parameters of the accretion disk and variability, the latent space of the SDE, the normalization of the light curve, and the uncertainty in the time-delay estimates. While $\boldsymbol{\eta}$ and $\boldsymbol{\hat{z}}$ are adjusted iteratively, the rest of the inferred parameters are produced directly each iteration. Each MLP takes as input the context, the mean and standard deviation of the input light curve, as well as the other previously predicted parameters of the network, while $\boldsymbol{\eta}$ and $\boldsymbol{\hat{z}}$ also use the output of the RIM network. To adjust $\boldsymbol{\eta}$ and $\boldsymbol{\hat{z}}$, we use an additional two-layer network, depending on the RIM network output and iteration number, that scales the updates by a factor dependent on the iteration count, allowing the network to gradually suppress changes as the RIM approaches convergence.

There are two MLPs in the latent SDE: the posterior drift function that decodes the context and latent vector, and the diffusion network that is applied element-wise to satisfy the diagonal noise~\citep{Torch_SDE}. The latent SDE solves the equation:
\begin{equation} \label{eq:latend_SDE}
\text{d}z(t) = \mu(t, z(t); \theta_\mu)\, \text{d}t + \sigma(t, z(t); \theta_\sigma)\, \text{d}W(t) \, ,
\end{equation}
where $\mu(t, z(t); \theta_\mu)$ is the drift term that determines the deterministic component of the latent dynamics with learnable parameters of the NN $\theta_\mu$, $\sigma(t, z(t); \theta_\sigma)$ is the diffusion term that represents the stochastic component of the latent dynamic with learnable NN parameters $\theta_\sigma$, $W(t)$ is a Wiener process (Brownian motion) that captures the random fluctuations over time, and $z(0) = \boldsymbol{\hat{z}}$ is the latent vector or initial condition of the SDE. The latent SDE consists of an Itô SDE solver using the Euler–Maruyama numerical approximation scheme broken up into 2,000 time intervals of 2.3 days each. 

We use a Gaussian mixture model to parameterize the posterior of the accretion disk and variability parameters with five multivariate Gaussians (discussed further in Section~\ref{sec:loss}). For the mixture coefficients, we use a softmax activation function to normalize them to probabilities. The variability parameters take as additional input the results from analyzing the reconstructed driving signal by conducting five linear fits to different pieces of the power spectrum and using the slope and $y$-intercept. We also use the mean, standard deviation, mean absolute deviation, total variation, and total square variation, as well as the results of the two latent SDE MLPs at $\boldsymbol{\hat{z}}$. To obtain a vector of accretion disk parameters $\boldsymbol{\hat{\eta}}$ from the posterior distribution, we sample from the posterior and take its mean. We use the mean times in the reconstructed transfer functions to obtain the relative time delays with respect to the reference $i$ band, and we assign uncertainties to the prediction. To normalize the light curve, we include three terms: a mean and standard deviation for each band's brightness in magnitude, as well as an additional bias term in the flux. The bias term is required since the host galaxy can contribute non-variable flux. We note that when a band is not observed, its mean brightness is set to the maximum limit of $27$ mag and standard deviation to $1$ mag to inform the ML model. 

Each MLP consists of four fully connected layers. The RNNs use tanh activation, transformers use GELU~\citep{hendrycks2023gaussianerrorlinearunits}, and fully connected layers use LeakyReLU~\citep{maas2013rectifier}. Whenever appropriate, we include residual skip connections~\citep{ResNet} and layer normalization~\citep{layer_norm} to enhance gradient flow, stabilize training, and improve model convergence. We use a hidden size of 256, transformer encoder size of 512, context size of 128, and latent size of 16. The transformer encoders have five layers, eight heads, and use a sinusoidal time embedding. 

\section{Robustness test} \label{sec:appendix_robustness}

\begin{figure}
    \centering
    \includegraphics[width=0.97\linewidth]{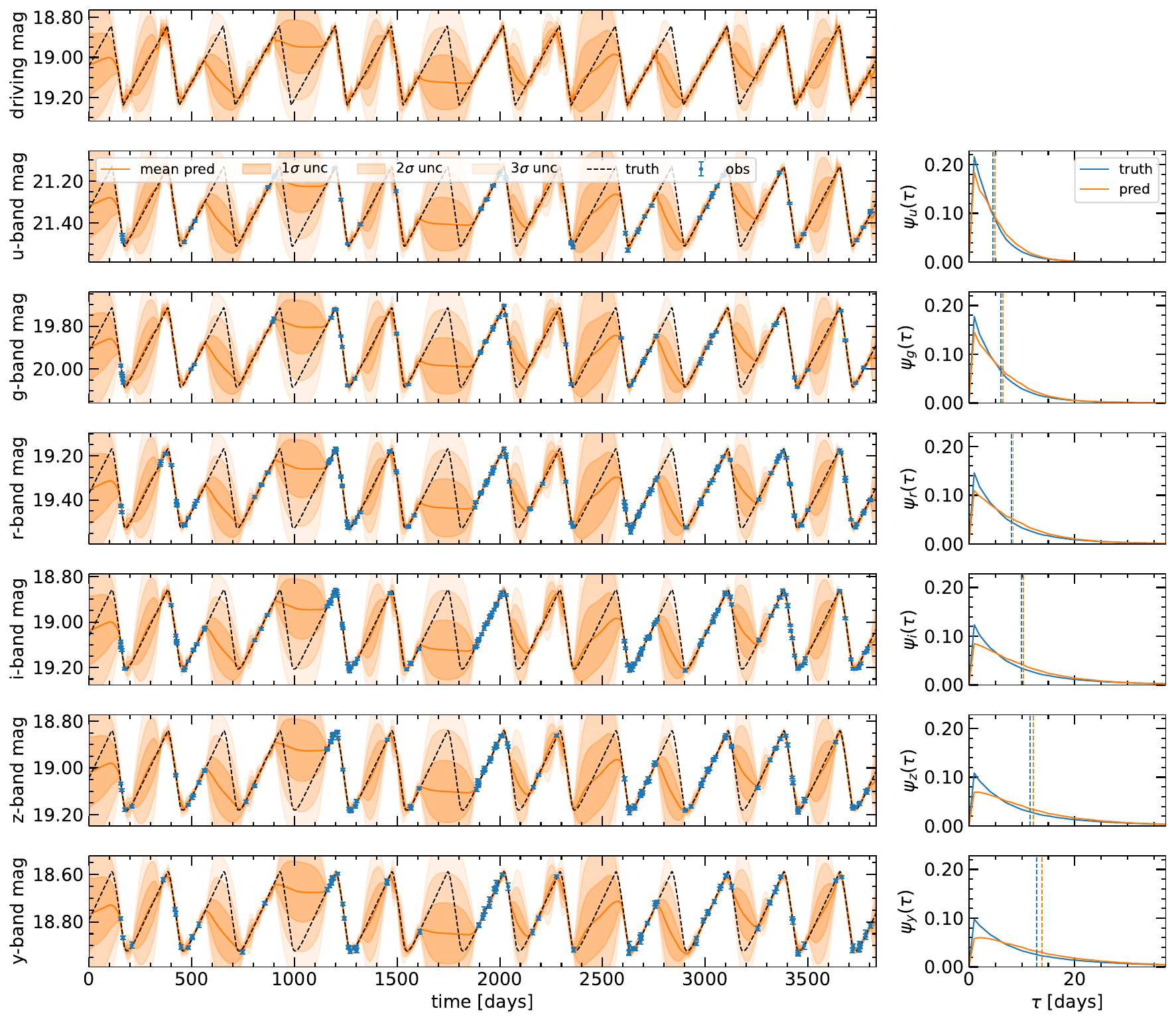}
    \caption{Same as Figure~\ref{fig:example_reconstruction} but applying our pretrained model to a test light curve with sawtooth driving signal.}
    \label{fig:example_sawtooth}
\end{figure}

\begin{figure}
    \centering
    \includegraphics[width=0.97\linewidth]{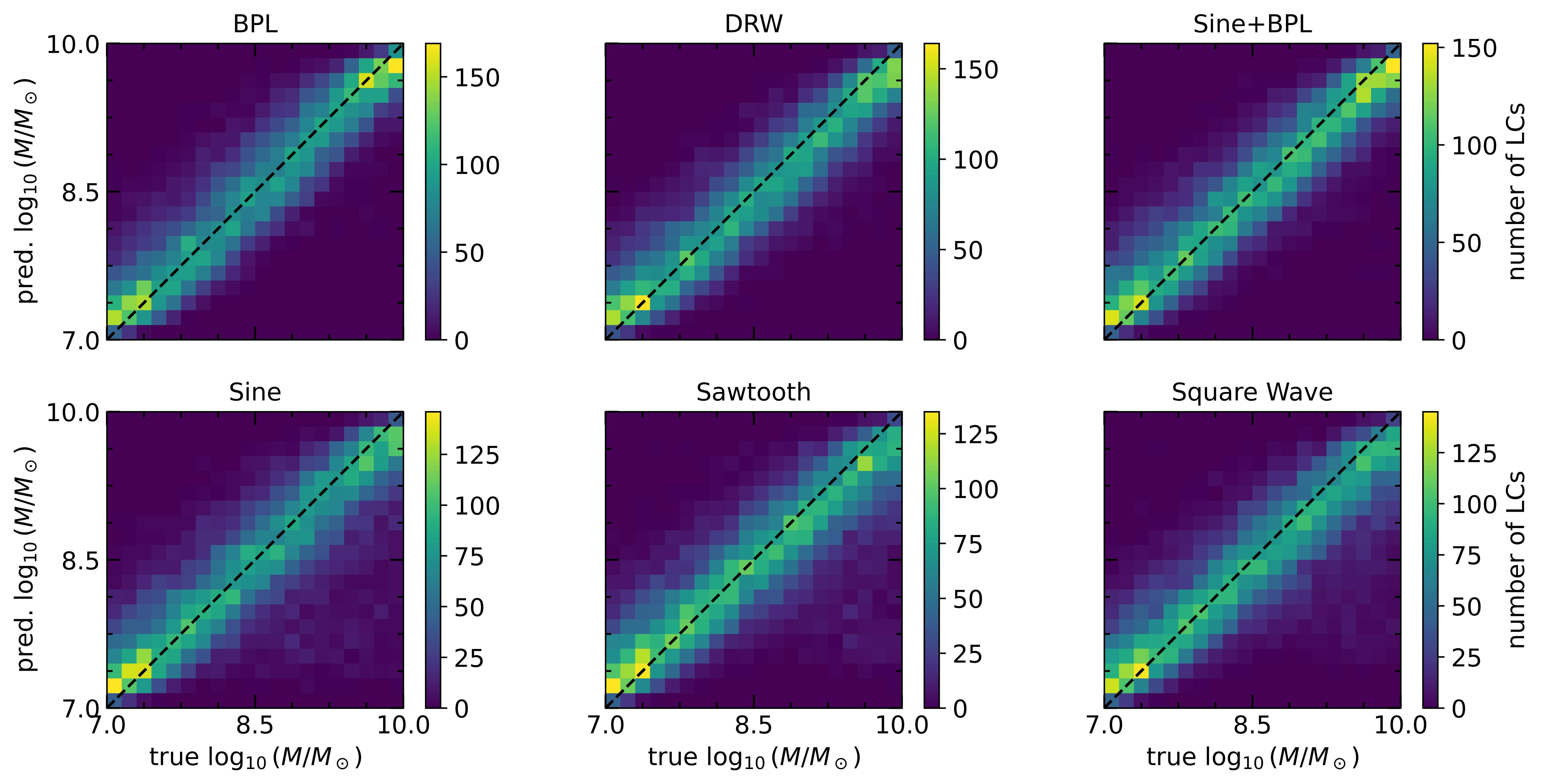}
    \caption{Same as Figure~\ref{fig:median_pred_vs_truth} for the black hole mass but applying our pretrained model to light curves with different out-of-distribution driving signals.}
    \label{fig:compare_mass_estimates}
\end{figure}

\begin{table}
\centering
\caption{RMSE normalized with the max$-$min of the uniform prior for each parameter. We note that predicting the mean of our uniform prior would yield a normalized RMSE of $1/\sqrt{12} \approx 28.87\%$.} 
 \begin{tabular}{c c c c c c c} 
Parameter & BPL & DRW & Sine+BPL & Sine & Sawtooth & Square wave \\
 \hline
$a$ & 28.80\% & 28.80\% & 28.80\% & 28.84\% & 28.82\% & 28.82\% \\ 
$\log_{10}(\lambda_{\text{Edd}})$ & 23.85\% & 23.96\% & 23.94\% & 24.35\% & 24.81\% & 24.74\% \\ 
$f_{\text{lamp}}$ & 28.92\% & 28.92\% & 28.93\% & 28.92\% & 28.93\% & 28.93\% \\ 
$(H-R_{\text{in}})/R_g$ & 28.78\% & 28.77\% & 28.77\% & 28.92\% & 28.89\% & 28.88\% \\ 
$\theta_{\rm inc}$ & 28.34\% & 28.43\% & 28.38\% & 29.64\% & 29.59\% & 29.46\% \\ 
$z$ & 6.44\% & 7.33\% & 6.90\% & 9.13\% & 9.59\% & 9.58\% \\ 
$\beta$ & 24.01\% & 24.00\% & 23.89\% & 24.61\% & 24.62\% & 24.64\% \\ 
$\log_{10}(M/M_\odot)$ & 11.08\% & 11.55\% & 11.46\% & 15.57\% & 15.54\% & 15.06\% \\ 
$\sigma/\text{mag}$ & 10.23\% & 12.88\% & 10.43\% & 17.92\% & 16.17\% & 17.05\% \\
$\log_{10}(\nu_b/\text{day}^{-1})$ & 18.94\% & - & - & - & - & - \\ 
$\alpha_L$ & 22.70\% & - & - & - & - & - \\ 
$\alpha_H-\alpha_L$ & 27.16\% & - & - & - & - & - \\  
\end{tabular}
\label{table:parameter_RMSE}
\end{table}

To test the robustness of our trained ML model to out-of-distribution variability, we compare its performance on test sets of light curves with BPL, DRW, BPL+sine, sine, sawtooth, and square wave driving signals (see Section~\ref{sec:light_curve_performance}). The BPL is our nominal test set and uses the same parameter range as our training set, given in Table~\ref{table:Parameter_table}. The DRW is the same as the BPL but with $\alpha_L = 0$ and $\alpha_H = 2$, which is outside the parameter range of our training set. We use a sine function with period $\log_{10}(T) \in [1, 3]$. For the sine+BPL, there is an additional parameter to determine the relative amplitude of the BPL, selected in the range of $[0.1, 2.0]$. The sawtooth has period $\log_{10}(T) \in [1, 3]$ with relative width between each saw edge chosen between $[0, 1]$. The square wave has period $\log_{10}(T) \in [1, 3]$ and relative width of each rectangular pulse chosen between $[0.1, 0.9]$. For each periodic signal, we select a random phase to shift the start time. We show an example reconstruction of a light curve with sawtooth driving variability in Figure~\ref{fig:example_sawtooth}. Despite the power spectrum of a sawtooth wave being very different than a BPL, our model still fits the observations and the signal is contained within the predicted $2\sigma$ credible interval. Since our model is trained using only stochastic signals, it will not recognize that this light curve is actually periodic but is still able to reconstruct the accretion disk parameters and time delays between wave bands.

To evaluate our model's ability to predict the accretion disk parameters for out-of-distribution light curves, we compare the median black hole mass prediction to the true values for each test set in Figure~\ref{fig:compare_mass_estimates}. We also give the normalized RMSE for each test set in Table~\ref{table:parameter_RMSE}. Our model can estimate each accretion disk parameter similarly to Figure~\ref{fig:median_pred_vs_truth}, although there are more outliers when predicting the redshift, temperature slope, and Eddington ratio. Furthermore, it can still infer the time delays between wave bands, as demonstrated in Figure~\ref{fig:example_sawtooth}.

\bibliography{bib}{}
\bibliographystyle{aasjournal}

\end{document}